\pdfoutput=1
\documentclass[12pt]{article}

\usepackage{amsmath,graphicx}
\usepackage{simplewick}
\usepackage{epsf}
\usepackage{graphicx,epsfig}
\usepackage{amsfonts}
\usepackage{amssymb}
\usepackage[nosort]{cite}
\usepackage{setspace}
\usepackage{caption}
\usepackage{color}

\def\bk{{\bf k}}

\def\CD{{\cal D}}

\def\CH{{\cal H}}
\def\CL{{\cal L}}
\def\CO{{\cal O}}

\def\tR{{\tilde R}}

\def\high{\vphantom{\Biggl(}\displaystyle}

\def\mpl{M_{\rm P}}
\def\mPl{M_{\rm P}}
\def\half{\frac{1}{2}}

\makeatletter
\renewcommand\section{\@startsection {section}{1}{\z@}%
                                 {-3.5ex \@plus -1ex \@minus -.2ex}%
                                   {2.3ex \@plus.2ex}%
                                   {\normalfont\large\bfseries}}
\renewcommand\subsection{\@startsection{subsection}{2}{\z@}%
                                   {-3.25ex\@plus -1ex \@minus -.2ex}%
                                     {1.5ex \@plus .2ex}%
                                     {\normalfont\bfseries}}
\renewcommand\subsubsection{\@startsection{subsubsection}{3}{\z@}%
                                   {-3.25ex\@plus -1ex \@minus -.2ex}%
                                     {1.5ex \@plus .2ex}%
                                     {\normalfont\itshape}}
\makeatother

\newcommand{\Letter}{
\setlength{\textwidth}{16.5cm}
   \setlength{\textheight}{22.6cm}
    \hoffset=-0.5in
\voffset=-2.1cm }

\Letter

\begin{document}
\newcommand{\be}{\begin{equation}}
\newcommand{\ee}{\end{equation}}
\newcommand{\bea}{\begin{eqnarray}}
\newcommand{\eea}{\end{eqnarray}}
\newcommand{\ba}{\begin{eqnarray}}
\newcommand{\ea}{\end{eqnarray}}
\newcommand{\barr}{\begin{array}}
\newcommand{\earr}{\end{array}}
\def\bal#1\eal{\begin{align}#1\end{align}}

\thispagestyle{empty}
\begin{flushright}
\end{flushright}

\vspace*{0.3in}
\begin{spacing}{1.1}

\begin{center}
{\large \bf Models of the Primordial Standard Clock}

\vspace*{0.3in} {Xingang Chen$^1$, Mohammad Hossein Namjoo$^1$ and Yi Wang$^2$}
\\[.3in]
{\em
$^1$Department of Physics, The University of Texas at Dallas, Richardson, TX 75083, USA \\
$^2$Center for Theoretical Cosmology,\\
Department of Applied Mathematics and Theoretical Physics, \\
University of Cambridge, Cambridge CB3 0WA, UK} \\[0.3in]

\end{center}

\begin{center}
{\bf
Abstract}
\end{center}
\noindent
Oscillating massive fields in the primordial universe can be used as Standard Clocks. The ticks of these oscillations induce features in the density perturbations, which directly record the time evolution of the scale factor of the primordial universe, thus if detected, provide a direct evidence for the inflation scenario or the alternatives. In this paper, we construct a full inflationary model of primordial Standard Clock and study its predictions on the density perturbations. This model provides a full realization of several key features proposed previously. We compare the theoretical predictions from inflation and alternative scenarios with the Planck 2013 temperature data on Cosmic Microwave Background (CMB), and identify a statistically marginal but interesting candidate. We discuss how future CMB temperature and polarization data, non-Gaussianity analysis and Large Scale Structure data may be used to further test or constrain the Standard Clock signals.

\vfill

\newpage
\setcounter{page}{1}

\tableofcontents

\newpage

\section{Introduction}
\setcounter{equation}{0}

Observationally distinguishing the inflation scenario from other possible alternative scenarios, as the origin of the Big Bang, remains an outstanding challenge for modern astrophysics and cosmology. The simplest inflation models \cite{Guth:1980zm,Linde:1981mu,Albrecht:1982wi,Starobinsky:1980te,Sato:1980yn,Mukhanov:1981xt,Hawking:1982cz,Starobinsky:1982ee,Guth:1982ec,Bardeen:1983qw} have received strong support from observational results on Cosmic Microwave Background (CMB) and Large Scale Structures (LSS) \cite{Komatsu:2010fb,Ade:2013zuv,Ade:2013uln}. In the meanwhile, there are also various attempts to construct alternative scenarios to explain the same observational results, see \cite{Brandenberger:2012zb,Ijjas:2013vea,Guth:2013sya,Linde:2014nna} for recent discussions. One of the reasons that such attempts are still possible is that there are only two parameters in the Standard Model of cosmology that are related to the primordial scenario, namely the amplitude and spectral index of the density perturbations. Therefore, an important research activity is to explore signatures beyond the Standard Model of cosmology, both in theories and most importantly in experiments.

While all observational signatures beyond the Standard Model are extremely valuable, not many of them can be used to {\em model-independently} distinguish inflation from the alternatives. Many signatures, if discovered, provide information that are helpful to distinguish different models within one scenario.

So far, there are two kinds of observational signals that are known to be capable of model-independently distinguishing the inflation scenario from the alternatives.

The first and well-known one is the primordial gravitational wave \cite{Grishchuk:1974ny,Starobinsky:1979ty,Rubakov:1982df}, namely the tensor mode, which may be detected in terms of the B-mode polarization in the CMB \cite{Seljak:1996ti,Seljak:1996gy,Kamionkowski:1996zd}. The tensor mode records the magnitude of the Hubble rate during the primordial epoch, therefore if detected distinguishes scenarios with fast-evolving scale factors, such as inflation, from scenarios with slowly-evolving scale factors, such as Ekpyrosis \cite{Khoury:2001wf}. However, phenomenologically, it is difficult to use the tensor mode to {\em model-independently} distinguish scenarios that all have fast-evolving scale factors and generate scale-invariant tensor modes. An example is the fast-expansion scenario (namely inflation) versus the fast-contraction scenario (such as the matter contraction \cite{Wands:1998yp,Finelli:2001sr}). Even in models with slowly-evolving scale factors, modified gravitational dynamics and/or non-vacuum initial condition for perturbations may lead to both nearly scale-invariant scalar and tensor power spectra (such as the string gas cosmology \cite{Brandenberger:1988aj, Nayeri:2005ck}).

The second one is the primordial Standard Clock signal \cite{Chen:2011zf,Chen:2011tu,Chen:2012ja,Chen:2014joa}, which may be detected as specific oscillatory features in the density perturbations. This signal is generated by the oscillatory massive fields in the primordial epoch. The classical oscillation pattern of massive fields in any time-dependent background is simple and can be regarded as a clock that generates standard ticks. These ticks imprint themselves as special oscillatory patterns in the density perturbations, which directly record the time evolution of the scale factor $a(t)$ of the primordial universe. The direct measurement of $a(t)$ provides a model-independent direct evidence for the inflation scenario or the alternatives.

In this paper, we study models of the primordial Standard Clock. There are two main objectives in this paper. One is CMB data analyses, another is theoretical model building. Firstly we compare the most important component of the Standard Clock signal, namely the clock signal, with the Planck 2013 CMB temperature data using Markov Chain Monte Carlo (MCMC). Secondly, we construct full models of Standard Clock in the inflation scenario.
We compute theoretical predictions of these full models, as a completion of the partial theoretical predictions used in the first part of the data analyses.
This enables us to make a full scale comparison between the theoretical prediction and CMB data.
While the statistical significance of the best-fit models from the MCMC analysis remains marginal, we show that there is an interesting candidate in the data which possesses the several key properties of the Standard Clock in the inflation scenario. This candidate can be further tested or constrained in the future by the CMB temperature and polarization data, non-Gaussianity analysis and Large Scale Structure data.

This paper is organized as follows. In Sec.~\ref{Sec:main_properties} we summarize the main model-independent properties of the primordial Standard Clock proposed previously.
The Standard Clock model prediction turns out to be a mixture of different types of features. We emphasize which are the most important properties that can be used to distinguish different primordial universe scenarios.
In Sec.~\ref{Sec:MCMC_clock}, we compare the clock signal alone with the Planck 2013 data and identify some interesting candidates. We compare the MCMC results between the inflation scenario and the alternatives. We also compare the clock signals with the pure sharp feature signal.
In Sec.~\ref{Sec:Full_Model}, we construct full models of Standard Clock in the inflation scenario. We motivate and construct a type of models in which the clock field is excited by tachyonic falling. We present full numerical results of the models in different parameter space, and discuss how to understand these results analytically. In Sec.~\ref{Sec:MCMC_full}, we compare the full prediction of the models with data. In Sec.~\ref{Sec:Polarization}, we use the best-fit model as an example to make predictions on the CMB polarization and the matter power spectrum of LSS, and discuss how primordial non-Gaussianities can be used to provide further tests. We conclude and discuss future directions in Sec.~\ref{Sec:Conclusion}. Appendices contain some detailed technical results used in the main text, including the full quadratic action in the perturbation theory and notes on numerical simulations.

\section{Main properties of Standard Clocks}
\label{Sec:main_properties}
\setcounter{equation}{0}

In this section, we summarize the main properties of the Standard Clock signals in the density perturbations.

To set up the phenomenological study, we use a simple power-law function to describe different kinds of primordial universe scenarios \cite{Chen:2011zf,Chen:2011tu},
\bea
a(t) \sim t^p ~,
\label{scale_factor}
\eea
where the time $t$ can run either from $0$ to $\infty$ or from $-\infty$ to $0$, and $p$ can be either positive or negative. This scale factor can describe different kinds of expanding or contracting universes.
The only requirement is that the quantum fluctuations of fields exit the horizon during the primordial epoch, in order to produce the acoustic oscillations in the CMB after these quantum fluctuations reenter the horizon during the Big Bang.
Given a value of $p$, this requirement fixes the range for $t$ and hence the choice between the expansion and contraction scenario. In the meanwhile, the magnitude of $p$ determines whether the scenario has fast or slowly-evolving scale factor.
Consequently, four qualitatively different scenarios can be characterized by the single fingerprint parameter $p$: $|p|>1$ corresponds to the fast-expansion scenario, namely inflation; $0< p \sim \CO(1)<1$ corresponds to the fast-contraction scenario, for example the matter contraction scenario; $0<p \ll 1$ corresponds to the slow-contraction scenario, for example the Ekpyrosis; $-1 \ll p <0$ corresponds to the slow-expansion scenario\footnote{Although in this case the scale factor is still acceleratedly expanding, the expansion speed is slow and $|\epsilon| \gg 1$.}.
The running directions of $t$ are as follows:
for $p>1$, $t$ runs from $0$ to $+\infty$; for all other $p$, $t$ runs from $-\infty$ to $0$. In terms of the conformal time $\tau$, $dt=a d\tau$, $\tau$ always runs from $-\infty$ to $0$. All the contraction scenarios need an extra bounce to match the Big Bang model.

The behavior of classical oscillations of massive field $\sigma$ in a given time-dependent background (\ref{scale_factor}) is simple and described by the equation of motion
\bea
\ddot \sigma + 3 H \dot\sigma + m_\sigma^2 \sigma =0 ~,
\label{massive_eom}
\eea
where $H = p/t$ is the Hubble parameter and $m_\sigma$ is the mass of the $\sigma$ field.

The time-dependent background (\ref{scale_factor}) has a horizon size above which the quantum fluctuations stop propagating. This horizon size is defined as
\bea
a(t) \int_t^{t_{\rm end}} \frac{dt}{a} =
|a\tau| = |t/(1-p)| ~.
\eea
For $p>1$, $t_{\rm end} = \infty$ because $t$ runs from 0 to $\infty$; for all other $p$, $t_{\rm end}=0$ because $t$ runs from $-\infty$ to 0. In terms of the conformal time, $\tau_{\rm end}$ is always 0.
This length scale corresponds to a critical mass scale $|(1-p)/t|$. When the mass $m_\sigma$ and time $t$ satisfy the relation $|m_\sigma t|> |1-p|$, we start to see the classical oscillations of the massive field following the solution of (\ref{massive_eom}).

This classical oscillation imprints its ticks in various background parameters, typically as some small oscillatory components. This in turn induces oscillatory components in the density perturbations, of which the patterns of ticks directly record the functional form of $a(t)$. Simple argument \cite{Chen:2011tu} shows qualitatively that, in the density perturbations, the oscillatory feature as a function of the wave-number $k$ is the inverse function of $a(t)$. Here we will summarize the quantitative results shortly.

As a comment, we note that in the leading order approximation, we have assumed that $m_\sigma$ is a constant and used the simple function (\ref{scale_factor}).
Sometimes it is natural to expect the various constants in this simple treatment to acquire some weak time-dependence.
These model-dependent variations within each scenario can be well tolerated by these approximations, because as we will see, for our purpose we only need to know the evolution within a timescale that is equal to or smaller than the horizon timescale, so any variations with a time scale that is larger than or comparable to the horizon timescale are not important.

The model-building realization of Standard Clocks can be very flexible, therefore, as emphasized in \cite{Chen:2011zf,Chen:2011tu}, it is important that we first extract the model-independent properties. In the following we summarize such properties. While all of them are interesting beyond-Standard-Model signals, we emphasize which properties can be robustly used to measure $a(t)$, which are less robust but nonetheless can be useful as auxiliary evidences, and which cannot be used to measure $a(t)$ and are universal for all scenarios.

Standard Clock models consist of two closely related processes. The first is the sharp feature process. This excites and initiates the classical oscillation of one or more massive fields. The second is the subsequent oscillation of the massive field, namely the clock field. Consequently, the full Standard Clock signal consists of two qualitatively different and intimately related signals, the sharp feature signal and the clock signal.

\begin{enumerate}

\item {\em The clock signal.}

    The clock signal is the most important part of the full Standard Clock signal. It is generated by the oscillation of the massive field, and manifest itself as a special type of oscillatory features in primordial power spectrum and non-Gaussianities.
    In the power spectrum the clock signal appears as a fractional correction to the leading-order nearly scale-invariant power spectrum, $\Delta P_\zeta/P_{\zeta 0}$; and in the non-Gaussianities it typically appears as a leading-order oscillatory component.
    Note that the running patterns of the clock signals in the power spectrum and non-Gaussianities are highly correlated. They can be uniformly written as
    \bea
    C \left(\frac{K}{k_r} \right)^\alpha
    \sin \left[ \frac{p^2}{1-p} \Omega \left( \frac{K}{k_r} \right)^{1/p} + \varphi \right] ~.
    \label{clock_signal}
    \eea
    Here $K$ is the wave-number: $K\equiv k_1+k_2=2k_1$ for power spectrum, $K\equiv k_1+k_2+k_3$ for bispectrum and so on;
    $\Omega$ is the ratio of the clock frequency $\omega$ induced by the massive field oscillation to the Hubble parameter $H_0 = H(t_0)$ ($t_0$ is the time of the sharp feature), so $\Omega$ is dimensionless; $k_r$ denotes the first resonant $K$-mode at $t_0$; $C$ is the amplitude;
    $\varphi$ is a constant phase, whose value varies for different correlation functions and models.
    For expansion scenarios, this formula applies for $K>k_r$; for contraction scenarios, this formula applies for $K<k_r$.

    As promised, we can now see quantitatively that the functional form inside the sinusoidal function in (\ref{clock_signal}) is exactly the inverse function of (\ref{scale_factor}). Thus if the amplitude of the clock signal is large enough in the density perturbations, measuring the value of $p$ provides the direct measurement of $a(t)$. For inflation, we only need to prove $|p|>1$.

    The value of $p$ determines the pattern of the ticks in the oscillatory feature, namely the relative distances between the consecutive zeros in (\ref{clock_signal}). This relative spacing is independent of the frequency $\Omega$, hence the mass of the clock field $m_\sigma$. The patterns of these zeros are imprints of the ticks of the massive field oscillation, and are insensitive to the nature of the couplings in detailed models. Because we are mostly interested in the relative spacing between the zeros, the patterns are also very likely to be unaffected by the physics of the yet-to-be-understood bounce in the contraction scenarios.

    Therefore, in this clock signal, these running patterns, determined by the fingerprint parameter $p$, are the most robust signatures for the primordial universe scenarios.

    The clock signal also has an overall envelop, parameterized by the factor in front of the sinusoidal function. The scale-dependence of this envelop also has some $p$-dependence. This dependence is less model-independent. For example, in power spectrum, if the coupling between the massive field $\sigma$ and density-perturbation-source-field is a direct coupling, then $\alpha = -3/2 +1/(2p)$ \cite{Chen:2014joa}; if the coupling is gravitational, then $\alpha = -3 + 5/(2p)$ \cite{Chen:2011zf}.
    The value of $\alpha$ can also be different in higher-point correlation functions.
    In addition, the direct coupling may be time-dependent, which would introduce some additional scale-dependence. On the other hand, the scale-dependence of this envelop can be nonetheless used as an auxiliary evidence for a scenario, because the oscillatory massive fields all have a fixed qualitative fate asymptotically in one scenario. For example, in inflation, all massive fields will be diluted; that is why no matter how complicated the detailed behavior is, the amplitude of this envelop decays away quickly towards large $K$ for inflation.

\item {\em The sharp feature signal.}

    The sharp feature that excites the massive field is also associated with a characteristic signal, which at the leading order can be qualitatively described by the following form,
    \bea
    \sim \cos(K/k_0+{\rm phase}) ~,
    \label{sharp_feature_signal}
    \eea
    where we have omitted a highly model-dependent envelop factor.
    Here $K$ is as defined in (\ref{clock_signal}), and the phase is model-dependent. The $k_0$ are the wave-number of the mode that is of the size of the horizon at the time of the sharp feature $t_0$. Like all sharp feature signals \cite{Chen:2010xka}, the same parameter $k_0$ approximately determines
    both the period of the sinusoidal running and the starting location of the sharp feature in the $K$-space.

    As we can see, the sharp feature signal, at least at this leading order, does not depend on the fingerprint parameter $p$. This is because the sharp feature only contains one click, hence no clock information. Variety of explicit examples of sharp features in the inflation scenario have been studied, which all demonstrate the behavior (\ref{sharp_feature_signal}) with model-dependent details. Still, we would like to emphasize here that our statement is stronger: this leading order behavior not only applies to the inflation scenario but also non-inflationary scenarios, therefore it cannot be used to distinguish inflation from the alternatives. Nonetheless this signal is an important part of the full signal, especially for the data analyses.

\begin{figure}[t]
  \centering
  \includegraphics[width=0.8\textwidth]{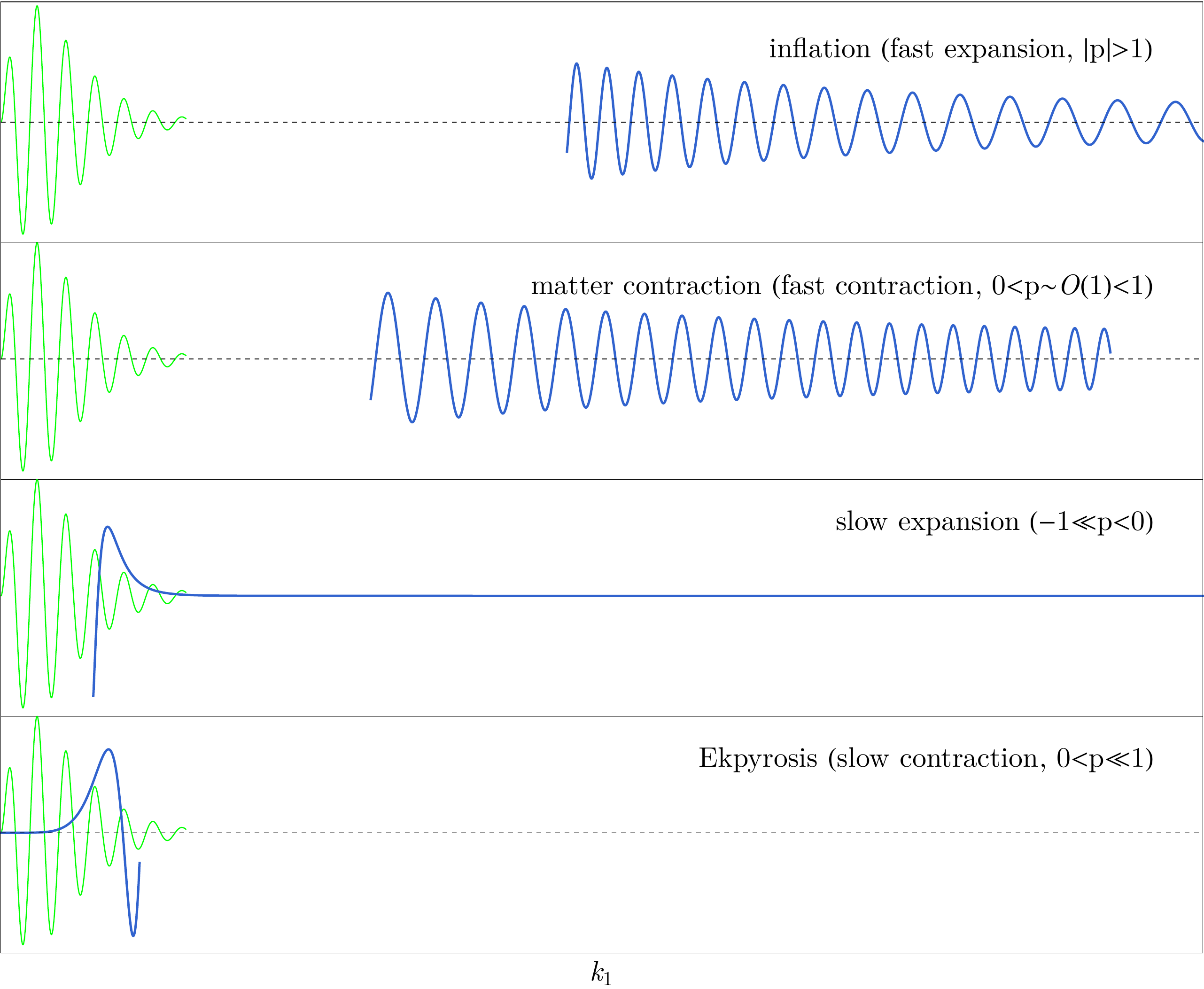}
\caption{A qualitative illustration of the full Standard Clock signals for different scenarios. Here we choose the same $\Omega$ for different scenarios ($\Omega$=120). Green/light lines represent the sharp feature signals; blue/dark lines represent the clock signals. The most important point in this illustration is the relative positions of the various ticks in the features.}
\label{Fig:scenarios1}
\end{figure}

\begin{figure}[t]
  \centering
  \includegraphics[width=0.8\textwidth]{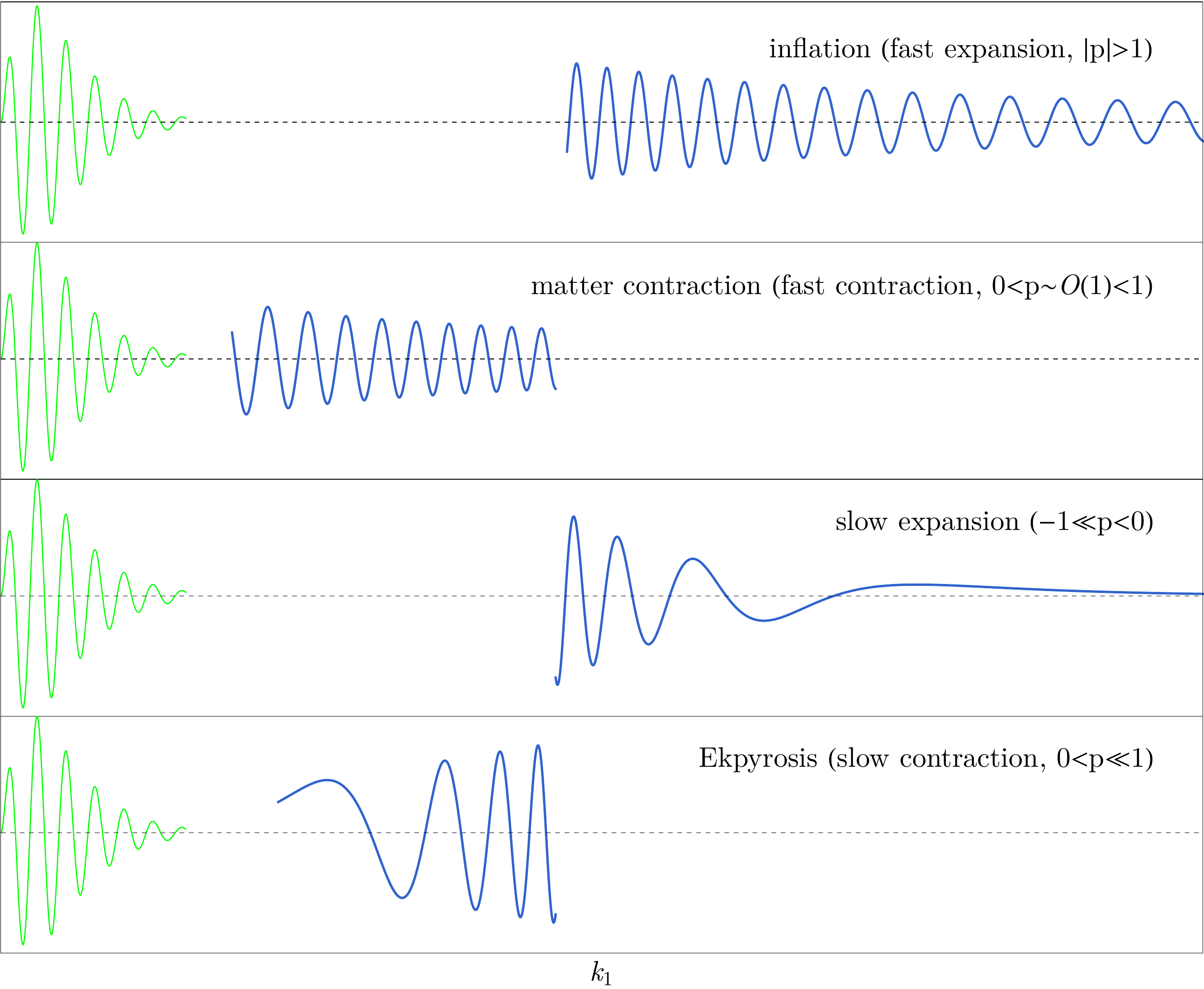}
\caption{The same as Fig.~\ref{Fig:scenarios1}, but here we choose the same $|\frac{p}{1-p}|\Omega$ for different scenarios, so that $k_r/k_0$ is the same for different scenarios.}
\label{Fig:scenarios2}
\end{figure}

\item {\em A relation between the clock signal and sharp feature signal.}

    The above two different kinds of signals show up at different locations in the $K$-space.
    This can be understood physically. The sharp feature perturbs the mode at the horizon, denoted as $k_0$, as well as neighbouring modes inside the horizon. The oscillating clock field, having a large frequency, excites modes that are well inside the horizon through the resonance mechanism, starting from $k_r$. So the ratio of these two scales are determined by the ratio of the frequency of the clock to the mass-scale of the horizon, which is scenario-dependent,
    \bea
    \frac{k_r}{k_0} = \frac{|p|}{|1-p|} \Omega ~.
    \label{k_relation}
    \eea

\end{enumerate}

The two signals and their relation, for different kinds of scenarios, are illustrated in Fig.~\ref{Fig:scenarios1} and \ref{Fig:scenarios2}. In Fig.~\ref{Fig:scenarios1}, we assume that the parameter $\Omega$ is the same for all scenarios. Later in the paper we will employ a more practical procedure in the data analysis, in which we try to look for the clock signals first in data.
This situation is closer to the case where the combination $|\frac{p}{1-p}|\Omega$ is held fixed for different scenarios. For this reason, we also present in Fig.~\ref{Fig:scenarios2} the same illustration but with fixed $|\frac{p}{1-p}|\Omega$. As we can see clearly in both figures, while the full signals spread over different scales and will be important for the full data analyses, the key information that we can use to distinguish different scenarios is the running behaviors, especially the patterns of the ticks, of the clock signals.

As a comment on the terminology, the special scale-dependent running behavior in the clock signal is a special case of the resonant running, due to the resonant mechanism \cite{Chen:2008wn} involved in the generation process; the special scale-dependent running behavior in the sharp feature signal is also called the sinusoidal running, which is a common feature of any sharp features \cite{Chen:2010xka}.

Finally to clarify several terminologies when we refer to different signals in this paper, by ``clock signal", we refer to (\ref{clock_signal}); by ``sharp feature signal", we refer to (\ref{sharp_feature_signal}); by ``full Standard Clock signal", we refer to the combined full signal that includes both the clock signal component and sharp feature signal component, an example of which will be worked out in Sec.~\ref{Sec:Full_Model}; and finally by ``pure sharp feature signal", we refer to the case in which the signal in the entire range of observable scales is the sharp feature signal.

\section{MCMC analysis on the clock signal}
\label{Sec:MCMC_clock}
\setcounter{equation}{0}

In this section, we use Planck (2013) + WMAP polarization to constrain the clock signal alone. The full signal, smoothly connecting the sharp feature signal and the clock signal, will be worked out and analyzed in Sec.~\ref{Sec:Full_Model} \& \ref{Sec:MCMC_full}. The public code \verb!CosmoMC! \cite{Lewis:2002ah} is used for the MCMC analysis.
We will present the best-fit models and the MCMC statistics, respectively, in the following two subsections. In these subsections, we discuss the results in terms of the following three kinds of comparisons.

\begin{itemize}

\item {\em Comparison between the clock signal and $\Lambda$CDM.}

In this comparison, we study whether the current data has any preference between the clock signals and the $\Lambda$CDM model. The fact that the Standard Clock models introduce several more parameters on top of the $\Lambda$CDM model should be taken into account when this comparison is made.

For this purpose, we make use of the Akaike's information criteria (AIC) \cite{AIC},
\begin{align}
  \mathrm{AIC} = 2k + \chi^2 ~,
\end{align}
where $k$ is the number of model parameters and $\chi^2$ is a result of MCMC. The quantity $\exp[({\rm AIC}_2 - {\rm AIC}_1)/2]$ measures the relative likelihood of model 1 over model 2.
In this paper, we only use this simple criteria to gain some rough ideas on the preference of models by data. As a result, our analysis will provide some interesting candidates. Any rigorous conclusions have to wait until more data become available.
We hope to compare the AIC with the Bayesian inference analysis in a future work.

\item {\em Comparison between inflation and alternative scenarios.}

To make this comparison, we perform MCMC using the clock signal (\ref{clock_signal}) with arbitrary $p$ value. The correction to the power spectrum can be written as
\begin{align}
  P_\zeta = P_{\zeta0} + \Delta P_\zeta ~,
\end{align}
\begin{align}
  \frac{\Delta P_\zeta}{P_{\zeta0}} = \begin{cases}
    0, & \text{$2k <  k_r$ for expansion scenarios},  \\
    0, & \text{$2k >  k_r$ for contraction scenarios}, \\
    \high{
    C \left( \frac{2k}{k_r}  \right)^{-\frac{3}{2}+\frac{1}{2p}}
    \sin \left[ \frac{p \Omega_\mathrm{eff}}{2} \left( \frac{2k}{k_r}  \right)^{\frac{1}{p} } + \phi \right]},  & \text{otherwise}.
  \end{cases}
  \label{clock_template}
\end{align}
In this section, for Standard Clock we concentrate on the clock signal only and have artificially truncated the signal to zero for $2k<k_r$ or $2k>k_r$ depending on the scenarios.
All the parameters in (\ref{clock_template}) are the same as in (\ref{clock_signal}) and we have defined $\Omega_{\rm eff} = 2\Omega p/(1-p)$.
The $\Omega_{\rm eff}$ can be either positive or negative, depending on the $p$ value. In the MCMC analyses, it is sufficient to restrict $\Omega_{\rm eff}$ to be positive because we keep the phase $\phi$ completely free.
The power in the envelop has been chosen to be the direct coupling case, $\alpha=-\frac{3}{2}+\frac{1}{2p}$.

As a reminder, ``expansion scenarios'' implies $p>1$ or $p<0$, while ``contraction scenarios'' implies $0<p<1$.
In terms of primordial universe models, it is useful to note that inflation takes place in the regime $p>1$, and $p<-1$ is the regime of super inflation. In this paper we consider inflation and super inflation together and refer them as inflation for short. Alternative to inflation has $-1 < p < 1$, which includes contraction and slow expansion. As demonstrated in \cite{Chen:2012ja} using mock data, the strength of the clock signals is in distinguishing the four qualitatively different kinds of scenarios by distinguishing the four different ranges for the parameter $p$. Once this leading degeneracy is broken, within the inflation scenario, the clock signal is not very sensitive to the detailed value of $p$, for example between $p=20$ and $p=30$, or between $p=-20$ and $p=20$, comparing to the other more conventional observables like the spectral index or scalar-to-tensor ratio. Therefore, for our purpose, to prove inflation we only need to show $|p|>1$.

In the figure legends, we refer to this comparison as ``inflation" versus ``alternatives".

\item {\em Comparison between the inflationary clock signal and pure sharp feature signal}.

It is also possible that sharp features in the models are not energetic enough to excite any massive fields, or the excited massive field is very massive and hence the clock signal starts appearing at much larger $k$-value in the momentum space. In both cases, we only see the pure sharp feature signal. As emphasized, a universal property of the sharp feature signal is its sinusoidal scale-dependence. We investigate how the data should be able to tell the different running behaviors between the clock signal and pure sharp feature signal.

In this comparison, for the clock signal we restrict to the inflation case. We take the exponential inflation limit, $p\rightarrow\infty$. Equation \eqref{clock_template} reduces to
\begin{align}
  \frac{\Delta P_\zeta}{P_{\zeta0}} = \begin{cases}
    0, & \text{if $2k<k_r$ ~,} \\
    \high{
    C \left( \frac{2k}{k_r}  \right)^{-\frac{3}{2}}
    \sin \left[ \frac{\Omega_\mathrm{eff}}{2} \log \left( \frac{2k}{k_r} \right) + \phi \right]},
    & \text{if $2k > k_r$ ~.}
  \end{cases}
  \label{clock_exp_inflation}
\end{align}

For the pure sharp feature signal we ignore the model-dependent envelop behavior, and simply use the template
\begin{align}
  \frac{\Delta P_\zeta}{P_{\zeta0}} =
    C \sin \left( \frac{k \Omega_\mathrm{eff}}{k_r} + \phi \right)
    ~.
    \label{sharp_sin_template}
\end{align}
In the full Standard Clock signal, the $2k<k_r$ part is essentially the same as the sharp feature, so to study the clock signal alone we only concentrate on the domain $2k>k_r$ in (\ref{clock_exp_inflation}). Therefore, to highlight the different running behaviors between the clock signal and pure sharp feature signal, in this section, we will also only concentrate on the $2k>k_r$ part for (\ref{sharp_sin_template}) and artificially set $0$ if $2k<k_r$. We have also replaced $k_0 = k_r/\Omega_{\rm eff}$ to facilitate this comparison.

In the figure legends, we refer to this comparison as ``inf clock" versus ``sin".

\end{itemize}

Finally, in this paper, we restrict our MCMC search for models with relatively low frequencies, namely we limit $\Omega_{\rm eff} \in [40,80]$. The high frequency search has the known problem of being easier to pick up statistical fluctuations with very small measure in the parameter space if we only compute the $\chi^2$. Nonetheless the high frequency search is also important and one can find examples in \cite{Ade:2013uln,Chen:2012ja} for different kinds of features. We hope to investigate this issue in the future.

\subsection{Best-fit parameters}
\label{Sec:Best-fit-parameters}

We present the best-fit parameters in Table~\ref{tab:best-chi2}. The $\chi^2$ is calculated by varying the model parameters, together with the 6 standard cosmological parameters and nuisance parameters.
The best-fit models are also plotted in terms of the corrections to the primordial signal and the CMB temperature anisotropy, respectively, in Fig.~\ref{fig:best}. The Planck data is also plotted for comparison purpose. The Planck data used in this paper is binned by width $\Delta\ell \approx 31$.\footnote{The data points in figures in \cite{Chen:2014joa} are binned by width $\Delta \ell \approx 25$, taken from Fig.~1 in Planck 2013 paper XVI \cite{Ade:2013zuv}. Note that there is a typo in the caption of that figure in the Planck paper: $\Delta \ell$ should be 25 instead of 31. In any case, the binning size only affects the visualization of data points in figures, but not the MCMC analyses in this paper which deals with the unbinned data.}
In the following, we discuss these results in more details by addressing the above-mentioned three comparisons.

\begin{table}[thbp]
  \begin{center}
    \begin{tabular}{ | c || c || c | c || c | c |} \hline
              & $\Lambda$CDM  & inflation & alternative & inf clock &  sin $(\dagger)$      \\ \hline\hline
$\log k_r$    &            & -2.23     & -2.47       & -2.34   &  -2.27    \\ \hline
$p$           &            & 105       & -0.691      &         &           \\ \hline
$\Omega_{eff}$      &            & 59.1      & 77.4        & 59.1    &  53.8     \\ \hline
$\phi$           &            & 2.15      & 3.02        & 5.40    &  6.12     \\ \hline
$C$           &            & 0.0576    & 0.107       & 0.0647  &  0.0450   \\ \hline\hline
$\chi^2_{\ell <50}$ & -6.760     & -6.676    & -5.791      & -6.926  &  -7.036   \\ \hline
$\chi^2_{\ell \geq 50}$ & 7795.276   & 7784.611  & 7780.185    & 7785.710&  7786.740 \\ \hline
$\chi^2_\mathrm{WP}$    & 2014.305   & 2014.296  & 2014.325    & 2014.291&  2014.274 \\ \hline
$\chi^2_\mathrm{total}$ & 9802.821   & 9792.231  & 9788.720    & 9793.075&  9793.978 \\ \hline
$\Delta\chi^2_\mathrm{total}$&            & -10.590    & -14.101 (*)     & -9.746   &  -8.843    \\ \hline
$\Delta$AIC        &            &  -0.590    &  -4.101 (*)      & -1.746   &  -0.843    \\ \hline
    \end{tabular}
  \end{center}
  \caption{\label{tab:best-chi2} \small Best-fit values for various models. The $\chi^2$ is calculated by the Planck likelihood code, which is not normalized (i.e.~only $\Delta\chi^2$ makes sense). The $\chi^2_{\rm WP}$ is from the WMAP polarization data. In searching for the best-fit models, not only the model parameters, but also the standard cosmological parameters $\{\Omega_bh^2, \Omega_ch^2, \theta, \tau, n_s, \log A\}$ are allowed to vary. Nevertheless the change in those standard cosmological parameters are minor and do not affect the fitting significantly.
  \\
  (*) Note that, although the alternative models have a slightly higher $\chi^2$ than inflation on average (see Fig.~\ref{fig:chi2}), the best-fit model listed here has a much lower $\chi^2$ than the best-fit of inflation. However, as we can notice from the lower-left panel in Fig.~\ref{fig:best} (and we have also checked from the $\chi^2$ analysis of the binned data), this best-fit alternative model becomes more significant because it is fitting a large dip near $\ell \sim 1700 - 2000$. According to version 3 of the Planck 2013 paper XVI (the last two paragraphs in Sec.~1) \cite{Ade:2013zuv}, this dip in data was later found to be due to a systematic error. While this systematic error has small effects on the conventional cosmological parameters, it is important for feature models like the ones being analyzed in this paper. The corrected data is not publicly available so far, so it will be interesting to revisit this issue after the next Planck data release.
  \\
  $(\dagger)$ We have identified another set of parameters for the ``sin" feature whose $\chi^2$ is smaller by 0.3 compared with the one presented. However, in that case $\log k_r = -3.37$, and $\Omega_\mathrm{eff}$ is comparable with the parameters listed here. As a result, that candidate has a much higher frequency, and is not comparable with the best-fit of the clock signal. Here we shall use the local best-fit with $\log k_r = -2.27$, instead of the global best-fit with $\log k_r = -3.37$, for the purpose of the comparison in Fig.~\ref{fig:best}.}
\end{table}

\newpage
\begin{itemize}

\item {\em Comparison between the clock signal and $\Lambda$CDM.}

We notice that the improvement of $\chi^2$ for different feature cases over the $\Lambda$CDM model is in general $\Delta\chi^2\sim 10$. Because these feature cases introduce 4 to 5 more parameters, according to the AIC, their statistical significance are all marginal and the data has no preference between them and $\Lambda$CDM. For example, comparing the clock signal with $\Lambda$CDM, the best-fit for inflation has $\Delta\chi^2 =-10.59$. Since the template (\ref{clock_template}) has 5 parameters, $\Delta {\rm AIC}=-0.59$. More properly when we restrict to the inflationary case, the clock signal has only 4 more parameters because the $p$-dependence is negligible as long as $p\gg 1$; so $\Delta {\rm AIC}\approx -2.59$ but still this does not make too much difference.

Nonetheless the various best-fit models provide interesting candidates to be investigated when more data is released, and for other types of analyses and experiments. We will discuss these aspects in Sec.~\ref{Sec:Polarization}.

\begin{figure}[t]
  \centering
  \begin{tabular}{cc}
  \includegraphics[width=0.45\textwidth]{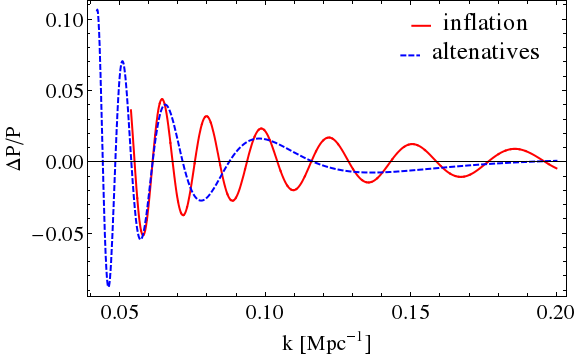}
  &
  \includegraphics[width=0.45\textwidth]{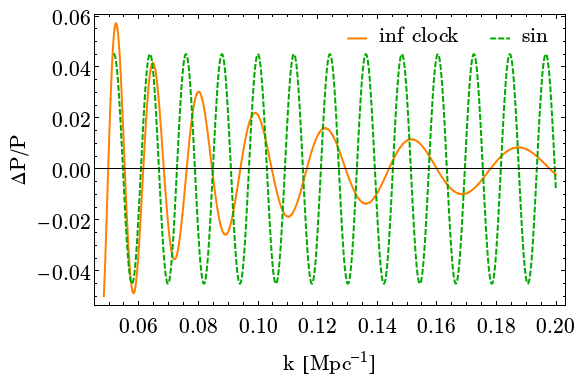}
  \\
  \\
  \includegraphics[width=0.45\textwidth]{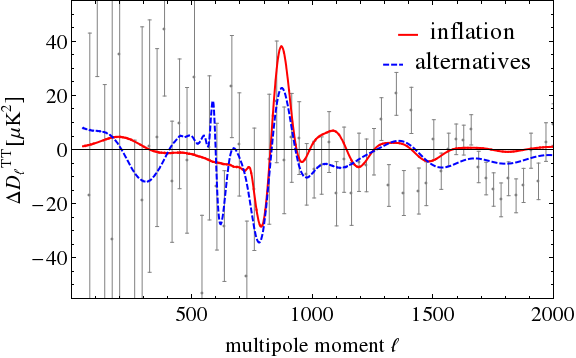}
  &
  \includegraphics[width=0.45\textwidth]{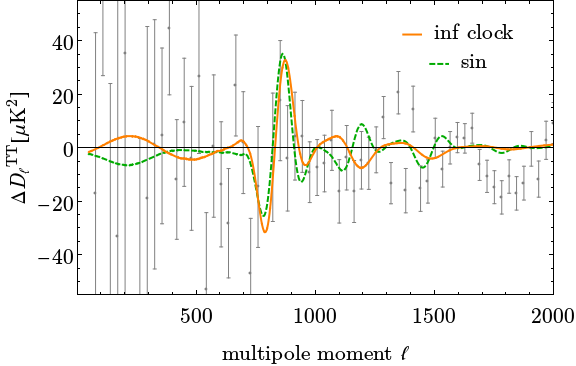}
  \end{tabular}
  \caption{\label{fig:best} \small
  The two plots in the upper panels are the corrections to the primordial power spectrum from the best-fit models, as functions of comoving wave number $k$.
  The two plots in the lower panels are the corresponding corrections to the CMB anisotropies from the best-fit models, as functions of multipole moment $\ell$. Here $D_\ell^{TT} \equiv  \ell(\ell+1)C_\ell^{TT} /(2\pi)$ and $\Delta D_\ell^{TT}$ is the correction to $D_\ell^{TT}$ of $\Lambda$CDM.
  Note that we have only plotted the clock signal component of the Standard Clock here, so in the lower two panels, some small oscillations in the lower $\ell$ regions ($\ell\lesssim 500$) can be ignored and are due to the artificial sharp truncation in Eq.~(\ref{clock_template}). For better comparison, the ``sin" signal is also truncated in this figure.}
\end{figure}

\item {\em Comparison between inflation and alternative scenarios.} (``inflation" versus ``alternatives").

Two best-fit models, one for inflation (with $p=105$) and another for an alternative model (with $p=-0.691$ hence does not belong to any known models in the literature) are presented in Table~\ref{tab:best-chi2}.
Naively this alternative-to-inflation best-fit appears to have a much better $\chi^2$ compared with either the inflationary clock signal or $\Lambda$CDM. However, as we look more closely at the plots of the actual fit in the upper-left and lower-left panels in Fig.~\ref{fig:best}, we can see that the improvement is due to a systematic error in the Planck data around $\ell \sim 1700-2000$. We discuss this in the caption of Table~\ref{tab:best-chi2}. Moreover, this best-fit model has a small measure in the parameter space. As we shall see in Sec.~\ref{sec:mcmc-statistics}, for typical choice of parameters (with flat prior) which fits data nicely, inflation is actually doing slightly better than the alternatives although the preference is marginal.

\item {\em Comparison between the inflationary clock signal and pure sharp feature}. (``inf clock" versus ``sin").

The best-fits for the inflationary clock signal (\ref{clock_exp_inflation}) and the pure sharp feature signal (\ref{sharp_sin_template}) are presented in the last two columns in Table~\ref{tab:best-chi2}, respectively. Both of them have 4 more parameters. Their statistical significance is comparable.

In fact this pure sharp feature candidate reproduces the low frequency candidate identified by the Planck Collaboration \cite{Ade:2013uln} using a sharp feature template with a much more complicated envelop. This candidate is also similarly reproduced in some subsequent data analyses \cite{Achucarro:2013cva,Achucarro:2014msa,Hu:2014hra} using a different sharp feature model.
As we can see, at the leading order the template (\ref{sharp_sin_template}) is sufficient for the model-independent purpose.

If we examine more closely where the best-fits pick up the statistical significance, we can find that both the inflationary clock template (\ref{clock_exp_inflation}) and the pure sharp feature template (\ref{sharp_sin_template}) improve the fits by fitting the wiggles around $\ell = 700 \sim 1000$. As we summarized in Sec.~\ref{Sec:main_properties} and will see more explicitly in the next two sections, for this candidate, below $\ell \lesssim 800$ the Standard Clock signal is dominated by the sharp feature signal, hence we expect to see the same oscillatory behavior as (\ref{sharp_sin_template}). The difference between the Standard Clock and the pure sharp feature signal starts to show up as $\ell >800$. We plot this difference part in the upper-right and lower-right panels in Fig.~\ref{fig:best}.
While the detailed envelop of the pure sharp feature is model-dependent, from the lower-right panel we can see that the most important difference between the two cases are the distinguished running patterns, resulting in different phases in the oscillations in $\Delta D_\ell^{\rm TT}$.
We can also see that
the current data is not precise enough to tell the difference between the two cases at any statistically significant level, although there are some very weak preference as we will see in the next subsection. Nonetheless this region belongs to the most important part of the full Standard Clock signal for this candidate, and in Sec.~\ref{Sec:Polarization} we discuss how the future data and analyses may help resolve the detailed scale-dependence in this region.

\end{itemize}

Another interesting aspect of the best-fit candidates is the following \cite{Chen:2014joa}. From Sec.~\ref{Sec:main_properties}, we know that for the Standard Clock there is a relation between the location of its clock signal and the location of its sharp feature signal. Let us use the best-fit model for inflation in Table~\ref{tab:best-chi2} as an example. This best-fit determines all the parameters in the Standard Clock model. In particular, the frequency is $\Omega =\Omega_{\rm eff}/2 \approx 30$, and the starting location of the clock signal is at $k=k_r/2 \approx 0.048 {\rm Mpc}^{-1}$ which is around $\ell \sim 660$. Using the relation (\ref{k_relation}), we can estimate the starting location of the sharp feature that is supposed to be associated with this clock signal candidate. In terms of the multipole space, this estimate gives $\ell \sim 22$. This matches a well-known sharp feature candidate in data at around $\ell=20\sim 30$, first identified by the WMAP collaboration \cite{Peiris:2003ff}. So the two well-separated features in the CMB, one at $\ell\sim 20$ and anther at $\ell\sim 700$, may well have a common origin from either a full Standard Clock signal or a pure sharp feature signal.

\subsection{MCMC statistics}
\label{sec:mcmc-statistics}

The best-fit models have tiny measure in the parameter space. Thus it is helpful to sample the parameter space through MCMC. In the MCMC statistics, we have fixed the 6 standard cosmological parameters. Note that, since the frequency range that we are interested in has some overlap with that of the acoustic peaks, the effect of the standard cosmological parameters is an interesting subject that we leave to future investigation.
We have provided flat prior for the model parameters $p$, $\Omega_\mathrm{eff}$, $\phi$ and $C$. On the other hand, flat prior is given to $\log (k_\mathrm{r} \mbox{Mpc})$ (which we shall refer to as $\log k_\mathrm{r}$) instead of $k_\mathrm{r}$ itself, considering that the scales vary exponentially during inflation or alternatives.

\begin{figure}[t]
  \centering
  \includegraphics[width=0.45\textwidth]{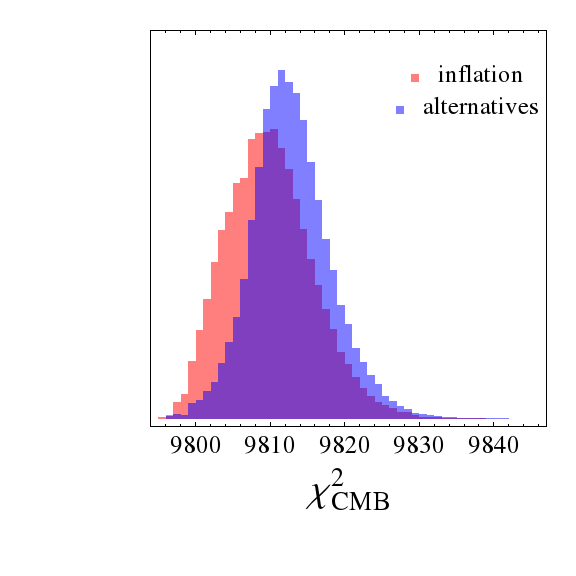}
  \includegraphics[width=0.45\textwidth]{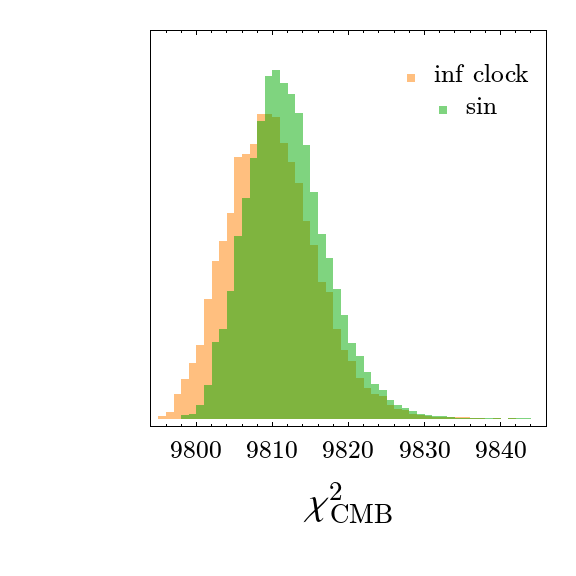}
  \caption{\label{fig:chi2} \small The distribution of $\chi^2$ in the MCMC chains. Inflation has lower $\chi^2$ on average compared with alternatives, with $\Delta\chi^2=
2.89$.
(We have also compared the distribution of inflation with three different types of alternative scenarios separately and observed very similar results as the figure in the left panel.)
Clock signal has lower $\chi^2$ on average compared with pure sharp feature signal, with $\Delta\chi^2=
1.91$. }
\end{figure}

\begin{figure}[htbp]
  \centering
  \includegraphics[width=1.0\textwidth]{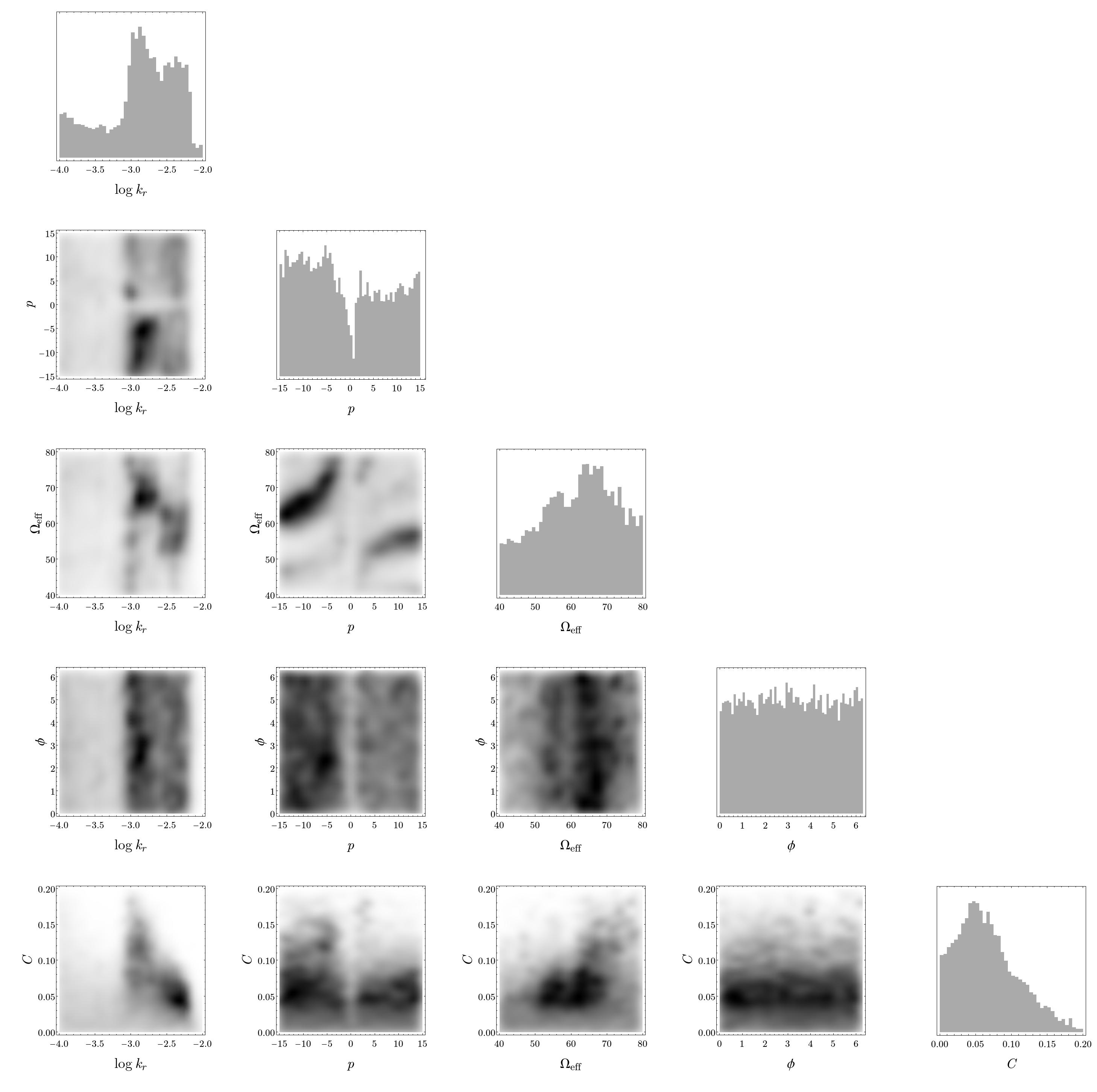}
  \caption{\label{fig:inf-alt} \small The triangle plot for model \eqref{clock_template}. Here the value of $p \in [-15,15]$ spans over inflation and alternatives. }
\end{figure}

\begin{figure}[htbp]
  \centering
  \includegraphics[width=1.0\textwidth]{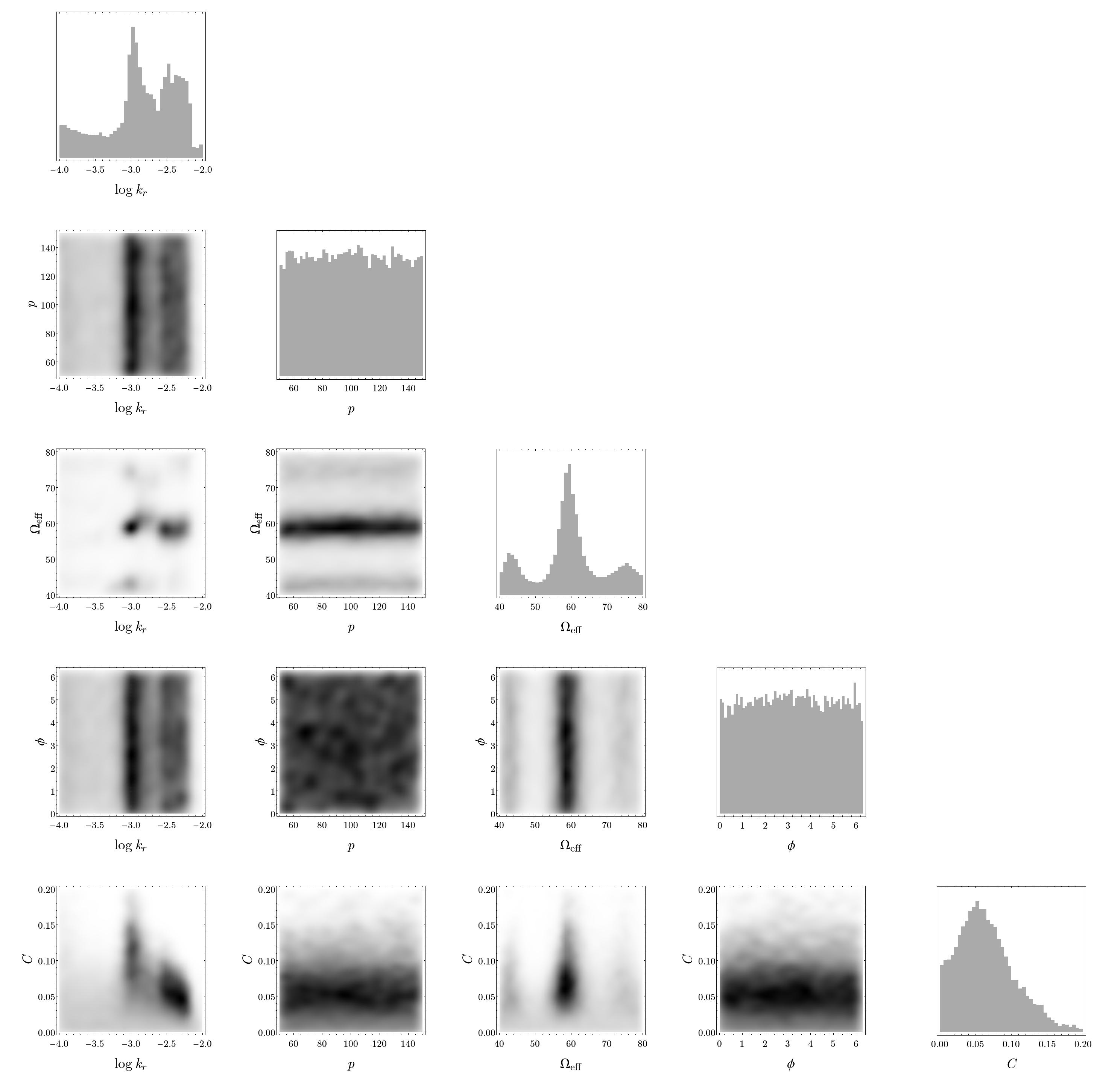}
  \caption{\label{fig:inf} \small The triangle plot for model \eqref{clock_template}. Here the value of $p \in [50,150]$ is chosen for inflation. }
\end{figure}

The results are presented in Fig.~\ref{fig:chi2}, \ref{fig:inf-alt} and \ref{fig:inf}. Below we discuss these results in terms of two comparisons.

\begin{itemize}

\item {\em Comparison between inflation and alternative scenarios.} (``inflation" versus ``alternatives").

We plot the $\chi^2$-distribution of inflation versus the alternatives in the left panel of Fig.~\ref{fig:chi2}, with all parameters marginalized. It is observed that given the Planck data, a typical model of inflation fits this feature candidate slightly better than the alternatives, with a difference $\Delta\chi^2 = 2.89$. Of course, the statistical significance is again marginal.

The MCMC samples are marginalized in the triangle plot Fig.~\ref{fig:inf-alt}, where $p \in [-15,15]$ to include both inflation and alternatives. It is interesting to note that the probability distribution function of $p$ has a dip at $p\sim 0$. This is consistent with
Fig.~\ref{fig:chi2}, showing that the alternatives-to-inflation is slightly less preferred in terms of fitting this feature candidate.
The $\Omega_{\rm eff}$-$p$ plot in Fig.~\ref{fig:inf-alt} reminds, albeit with much lower resolution, of the last plot in Fig.~13 in Ref.~\cite{Chen:2012ja} where the forecast for such analyses was made. As we emphasized, the clock signal is not very sensitive to the detailed value of $p$ within the inflation model, but is capable of showing $|p|>1$ once an inflationary clock signal is present above certain threshold in data.

To study the regime of inflation more closely, we zoom in to $p \in [50, 150]$. In this regime, the preference patterns in the samples are sharper. A peak at $\Omega_\mathrm{eff} \simeq 60$ is identified, where $\log k_\mathrm{r}$ is likely to be either $-2.4$ or $-3.0$. The amplitude $C$ has a preference at $C \simeq 0.05$. While the posterior distribution is quite flat for both $p$ and $\phi$.

\item {\em Comparison between the inflationary clock signal and pure sharp feature}. (``inf clock" versus ``sin").

We also plot the $\chi^2$-distribution of the inflationary clock signal versus the pure sharp feature signal in the right panel of Fig.~\ref{fig:chi2}.
A typical clock signal fits data slightly better than a pure sharp feature signal, by a statistically marginal $\Delta\chi^2=1.91$. This small difference is due to their different oscillatory phases in the $\ell>800$ region, demonstrated by the best-fit examples in the lower-right panel of Fig.~\ref{fig:best}. This region is potentially the most informative part of the signals for this candidate.

\end{itemize}

\section{Full models of Standard Clock in inflation}
\label{Sec:Full_Model}
\setcounter{equation}{0}

In Sec.~\ref{Sec:MCMC_clock} we have performed the MCMC analyses on the clock signal component of the full Standard Clock signal.
Although the clock signal and its index $p$ are the most important information we would like to search in data, the sharp feature signal is also an integral part of the full signal. Its existence, albeit with model-dependent details, is necessary for the full Standard Clock process; and its presence in data would provide an additional confirmation of the whole picture.
It is therefore important to construct full models of Standard Clock and supply the full details that link the several universal properties proposed in \cite{Chen:2011zf,Chen:2011tu} and summarized in Sec.~\ref{Sec:main_properties}.

In this paper we construct full models of the Standard Clock for the inflation scenario.
Full models for the alternative-to-inflation scenarios are also important research subjects. Similarly they help us to understand the detailed links between the two types of signals in those scenarios. We leave this for future investigation.

\subsection{Model-building requirements}

As emphasized in Ref.~\cite{Chen:2011zf,Chen:2011tu}, there are many possible ways that massive fields can get excited classically, and ways that these clock fields can be coupled to the density-perturbation-source-field.
Many models in the literature may well have already contained the necessary ingredients even though their roles as Standard Clock models have not been realized. Generally speaking, there are only two simple requirements for an inflation model to be qualified as a model of Standard Clock:

\begin{itemize}
\item
In the period of the last, observable, 60-efolds of inflation, we need at least two stages of inflation, which are connected through some kind of sharp feature. This requirement ensures that the period of the oscillation in the clock signal (\ref{clock_signal}) is large enough in the multipole space\footnote{Namely, the period $\Delta\ell >2\pi$ in the strongest part of the signal. If the sharp feature is encountered at or before the beginning of the observable 60-efolds of inflation, the period of the initial clock signal will be $\Delta\ell \le 2\pi$.} to be resolved by experiments in principle.

\item
The sharp feature that connects the two stages of inflation should excite a massive field which starts to oscillate classically. This massive field is the clock field.
\end{itemize}

These two ingredients also determine the key observable properties for the inflationary case as summarized in Sec.~\ref{Sec:main_properties}.
Nonetheless, once these two requirements are satisfied, model-dependent details are also important. Even though they do not change the key properties in Sec.~\ref{Sec:main_properties},
they affect the absolute and/or relative magnitudes of the two types of the signals. These details matter in data analyses, and sometimes can even become one of the crucial points. For this reason, in the rest of the paper, we set out to construct a full model realizing the key proposals of the Standard Clock that we have summarized so far. We work out its consequence rigorously in full details and compare the predictions with the Planck data.

\subsection{Model Lagrangian}

In previous studies, the examples commonly used to illustrate the points are models with sharp bending trajectory \cite{Chen:2011zf,Chen:2011tu,Gao:2013ota,Noumi:2013cfa,Saito:2012pd,Kobayashi:2012kc,Battefeld:2013xka,Mizuno:2014jja,Battefeld:2014aea}. These models seem to give much stronger sharp feature signal than the clock signal in the power spectrum, due to either the cancellation between different resonant contributions to the clock signal, or some relatively large spike in the sharp feature signal. Although this observation is still limited to the simplest examples that have been studied so far and other variations are worth to be explored, in this paper we pursue a different type of models \cite{Chen:2014joa} because this type of models have more appealing motivation from the point of view of UV-completion model-building. In addition we will also see some nice phenomenological aspects when they are compared with data. In these models, the clock field starts to oscillate simply because it falls into a potential well.

The model-building motivation for this type of models is as follows. It is well-known that, for slow-roll inflation, some kind of fine-tuning is needed to make the potential flat enough to satisfy the slow-roll conditions. There may be many fields rolling at the beginning of inflation, but only very few of them finds the flat potential and eventually drives the 60 e-folds of inflation. Most of the fields fall into some potential wells earlier and settle down at the bottom of those potentials. The curvature at the bottom of those potentials is larger than the Hubble parameter, so the fields oscillate for a few efolds before completely settling down. These fields are natural candidates for the clock field.

The Lagrangian of this model is the following,
\bea
\CL =
-\frac{1}{2} (\tR + \sigma)^2 g^{\mu\nu} \partial_\mu \theta \partial_\nu \theta - V_{\rm sr}(\theta)
-\frac{1}{2} g^{\mu\nu} \partial_\mu \sigma \partial_\nu \sigma - V_\sigma(\sigma) ~,
\label{model_Lagrangian}
\eea
where the potential
\bea
V_\sigma = V_{\sigma 0} \left[ 1- \exp(-\sigma^2/\sigma_f^2) \right]
\label{V_sigma}
\eea
models the potential dip for the clock field $\sigma$. The potential $V_{\rm sr}$ is any slow-roll potential. We will examine different slow-roll potentials in the following subsections.

\begin{figure}[t]
  \centering
  \includegraphics[width=0.7\textwidth]{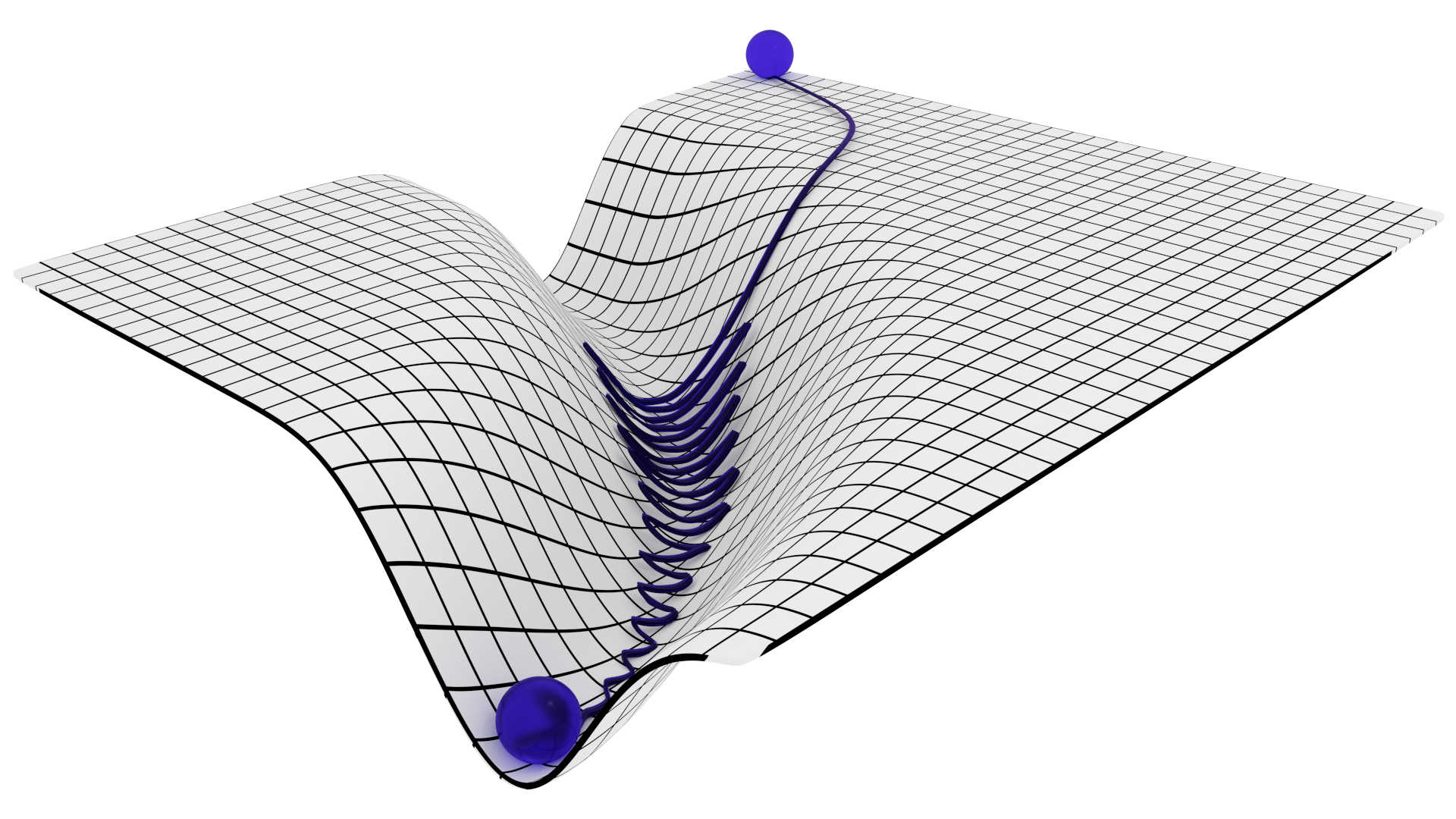}
  \caption{\label{fig:potential} \small An illustration of the model (\ref{model_Lagrangian}).
  The model starts as a two-field inflation model.
  The inflaton rolls on a plateau for a few efolds before falling into a potential valley. The massive field is excited and oscillating. Eventually the model settles down to the 2nd stage of inflation as an effective single-field slow-roll model.}
\end{figure}

This Lagrangian is the same as that of the quasi-single-field inflation model \cite{Chen:2009we,Chen:2009zp} with a large mass term \cite{Chen:2012ge,Pi:2012gf}. At late time, this model approaches to an effective single field inflation model, in which the inflaton field $\theta$ rolls along a slightly curved trajectory (with radius $\tR$) determined by the valley of the potential. The orthogonal direction is lifted by the potential (\ref{V_sigma}) with a large mass $m_\sigma \gg H$. This late-time behavior should be a very generic description of any low-energy effective theory of single-field slow-roll inflation models. In the current model, an additional ingredient is the inclusion of an earlier phase describing how the inflaton settles down at the bottom of the potential valley. For this purpose we added a plateau to the potential of the $\sigma$-field, as described in (\ref{V_sigma}). We initially put the inflaton field somewhere on this plateau. As mentioned, the generic shape of such a plateau should not be flat enough to support all 60-efolds of inflation; so after a few efolds, the inflaton falls into the bottom of the potential dip described by (\ref{V_sigma}), oscillates for a while, and settles down to an effective single field inflation model.
See Fig.~\ref{fig:potential} for an illustration.
This is the model that we shall study both in terms of the background and the density perturbations.

To have a complete understanding of the implications of this model on the observables, we note two important aspects in our following analysis that are distinguished from those in simple inflation models. Firstly, the Standard Clock originates from the background and as mentioned there are a variety of model-building possibilities. Although sharing several key properties, they give rise to different full signals that are important to data analyses. Therefore it is worth to investigate different backgrounds and the data may even have the ability to distinguish them. Secondly, for simple inflation models, the leading observable effects can often be captured by perturbing the matter sector alone while ignoring the perturbations in the gravity sector. This simple approach is sometimes referred to as the inflaton approximation or decoupling approximation. However as we will see, to have a complete understanding of this model in different parameter space, the perturbative terms coming from the gravity sector become important.

\subsection{Background}

\begin{figure}[t]
  \centering
  \includegraphics[width=0.45\textwidth]{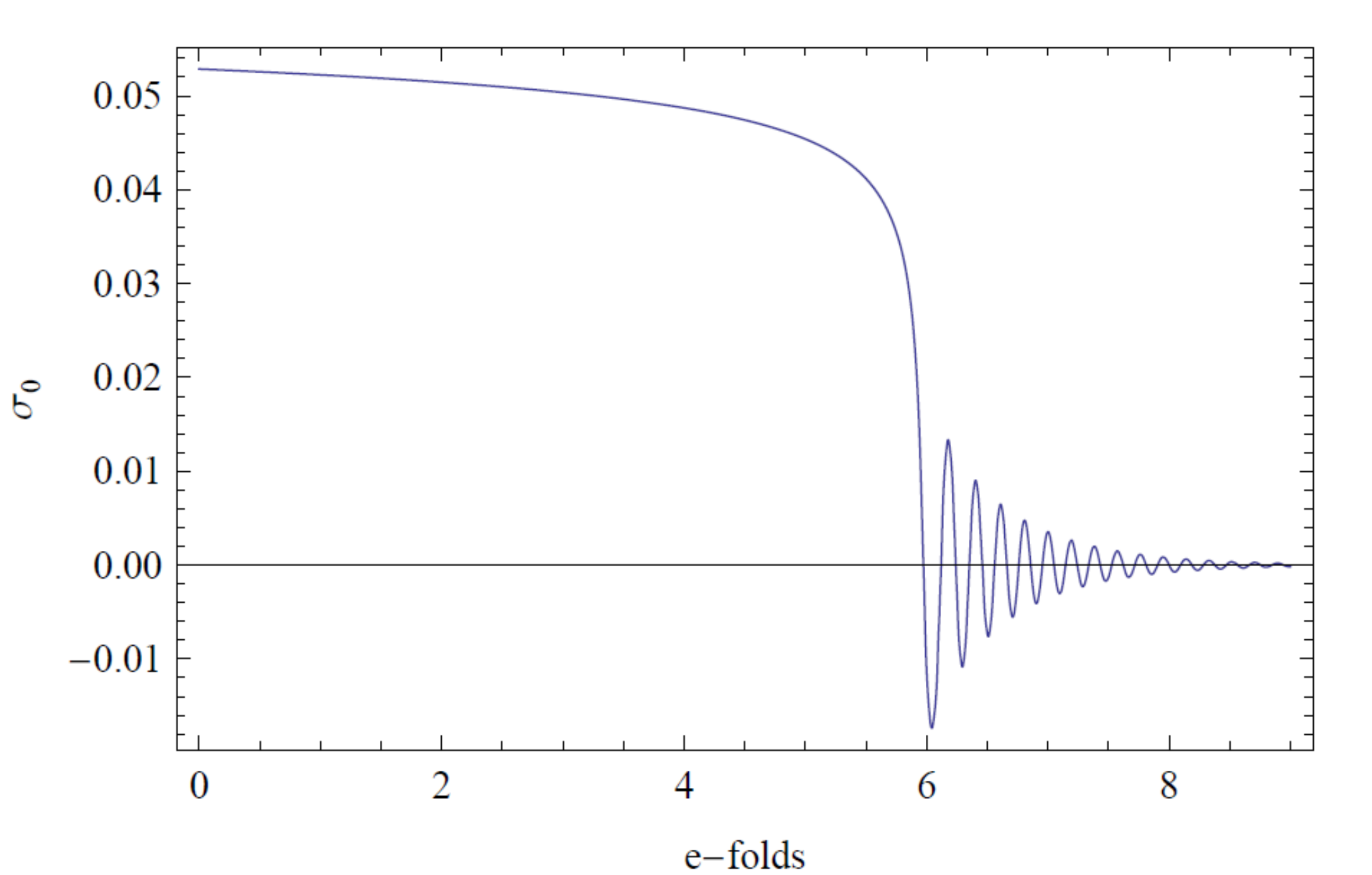}
  \includegraphics[width=0.45\textwidth]{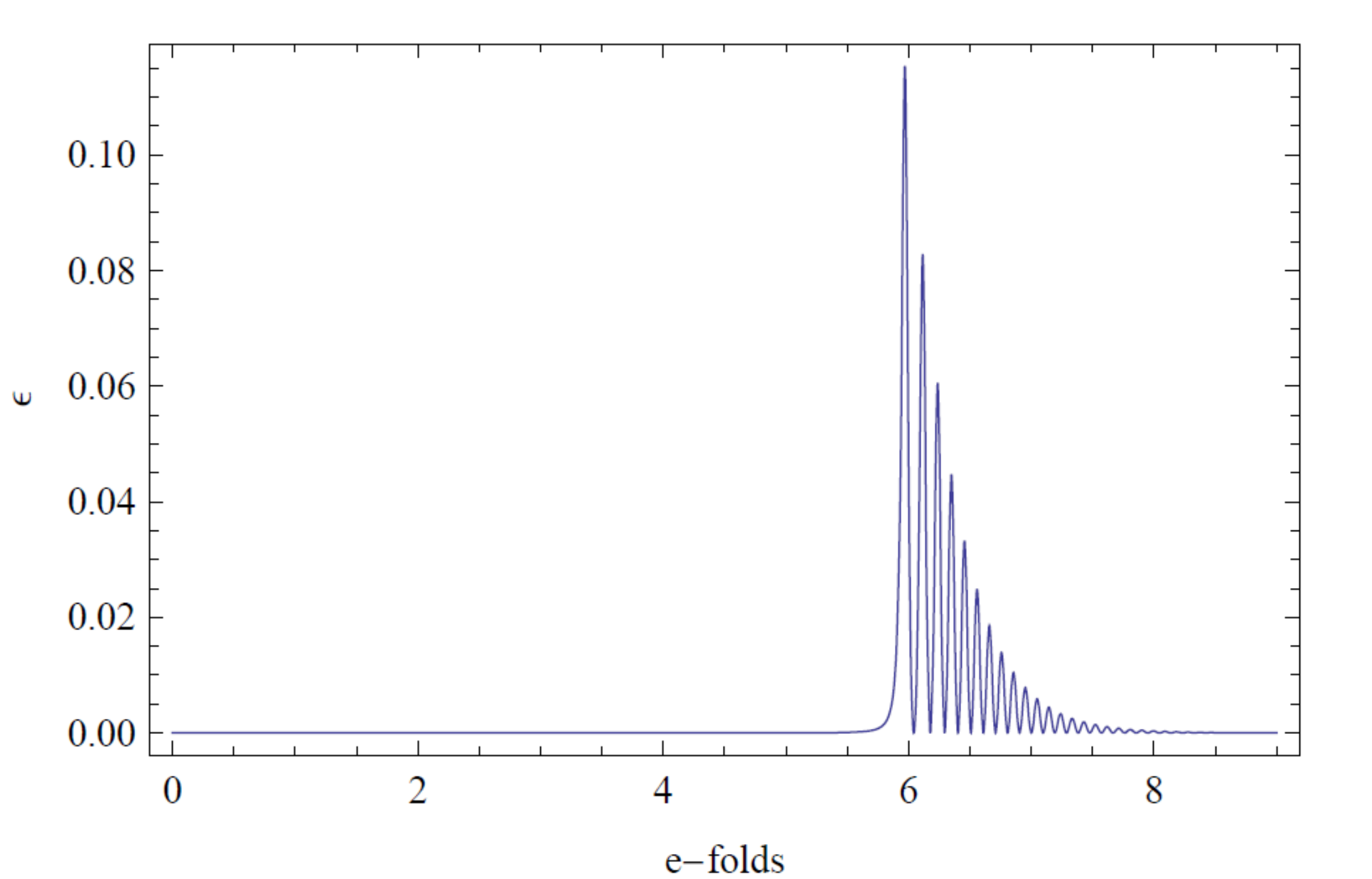}
  \caption{\label{Fig:bkgd_figs} \small An example of background evolution of the clock field $\sigma_0$ and the slow-roll parameter $\epsilon$. In this example, the attractor value of $\epsilon$ is $10^{-6}$. The model parameters in this example are $V_{\rm inf}=5.33\times 10^{-13}$, $V_{\sigma 0}=2.66\times 10^{-14}$, $\sigma_f = 1.64\times 10^{-2}$, $\tR=2.05$, $\beta=1.55\times 10^{-15}$. Initially we put the inflaton at the potential plateau $\theta_{\rm initial} =0$, $\sigma_{\rm initial}=8.33\times 10^{-2}$ so that the clock starts oscillating at $N \approx 6$ efolds. In this paper, we set $\mpl=1$.}
\end{figure}

We start with the background equations of motion,
\begin{align}
& 3\mPl^2 H^2 = \dfrac12 ( \tilde{R}+\sigma _0 )^2 \, \dot\theta _0^2+\dfrac12\dot\sigma _0^2+V_{\rm sr}+V_{\sigma }  ~,
\\
& (\tR + \sigma_0)^2 \ddot\theta_0 + 3 H (\tR+\sigma_0)^2 \dot\theta_0
+2 (\tR+\sigma_0) \dot\sigma_0 \dot\theta_0 + V'_{\rm sr} =0  ~,
\\
& \ddot\sigma_0 + 3H\dot\sigma_0 + V'_\sigma - (\tR+\sigma_0)\dot\theta_0^2
=0  ~.
\end{align}
We initially place the inflaton on the plateau of the $V_\sigma$ potential, and let it roll for a few efolds until it drops down towards the dip and oscillate. The oscillation frequency is determined by the mass term near the bottom of the potential $V_\sigma$, $m_\sigma \approx \sqrt{2 V_{\sigma 0}}/\sigma_f$, and the evolution of the oscillation amplitude is typical of the massive field, $\sim a^{-3/2}$. The full evolution can be solved numerically according to the model prescriptions given above. In Fig.~\ref{Fig:bkgd_figs}, we give an example of the background evolution of the $\sigma$-field and the slow-roll parameter. The slow-roll potential used in this example is a small field example
\bea
V_{\rm sr}(\theta) = V_{\rm inf} - \beta \theta ~,
\label{V_smallfield}
\eea
in which $V_{\rm inf}$ provides the dominant potential energy driving inflation. As we can see, when the clock field $\sigma$ oscillates, the slow-roll parameter $\epsilon$ receives a transient burst of oscillations with an amplitude orders-of-magnitude larger than its attractor value. Very similar behaviors are seen in different parameter space and in both small and large field inflation models. As we shall see, it is interesting how large can certain oscillatory behaviors in inflation models still be consistent with the current data.

\subsection{Perturbations}

In this paper, we restrict the study of the perturbation theory to the tree-level two-point correlation function.
For this purpose, we perturb all fields around their background values and expand the Lagrangian to the second order in perturbations. The details can be found in Appendix \ref{App:quadratic_action} and the following is the main result,
\bea
\CL_2 &\approx&
\frac{a^3}{2} (\tR + \sigma_0)^2
\left[ \dot{\delta\theta}^2 - \frac{1}{a^2} (\partial_i \delta\theta)^2 \right]
+ \frac{a^3}{2} \dot{\delta\sigma}^2 - \frac{a}{2} (\partial_i \delta\sigma)^2
\nonumber \\
&-& \frac{a^3}{2}
\left[V''_{sr}
-(\tR + \sigma_0)^4 \dot \theta_0^2 (\epsilon-3) +\dfrac{2}{H}(\tR + \sigma_0)^2 \dot \theta V'_{sr}
\right]
 \delta\theta^2
\nonumber \\
&-& \frac{a^3}{2}
V''_\sigma \delta\sigma^2
\nonumber \\
&+& \frac{a^3}{H} (\sigma_0+\tR) \dot\theta_0
\left( 2H - (\sigma_0+\tR) \dot\sigma_0 \right)
\delta\sigma \dot{\delta\theta}
\nonumber \\
&-& \frac{a^3}{H} (\sigma_0+\tR)^2 \dot\theta_0 \dot\sigma_0
\dot{\delta\sigma} \delta\theta ~.
\label{L2_leading}
\eea
Although in numerical calculations we will keep all terms in (\ref{L2_full}), for better analytical understanding we only listed the most important terms above. Notice that (\ref{L2_leading}) still includes terms from both the matter and gravity sectors. When estimating the order-of-magnitude of each term to determine its importance, we used the approximation that each time derivative on the background parameters (such as $\sigma_0$, $\dot\theta_0$, $\epsilon$) and the perturbations (such as $\delta\theta$, $\delta\sigma$) brings out a factor of $\omega\sim m_\sigma$, since the dominant behaviors of these parameters at the feature are oscillations with frequency $\sim \omega$. This estimation applies not only to the massive field oscillation process, but also the sharp feature transition process, because the sharpness of the sharp feature is determined by the curvature of the potential, which in turn is determined by the same mass term $m_\sigma$ for a reasonably smooth potential. The details of this estimation can be found in Appendix \ref{App:quadratic_action}.

It is straightforward to write down the equations of motion for the two fields $\delta\theta_\bk$ and $\delta\sigma_\bk$ in the momentum space.
We set the initial conditions for both fields to be in the Bunch-Davies vacua, namely,
\bea
(\tR+\sigma_0)\delta\theta_\bk~,~\delta\sigma_\bk
\to
\frac{H}{\sqrt{2k}} \tau e^{-ik\tau} ~,
\label{BD_vacuum}
\eea
where $\tau$ is the conformal time.
We set this initial condition when modes are at least a factor of $\omega/H$ deeper within the horizon, because we will be interested in certain subhorizon modes at the time of sharp feature.
The complete solution of the equations of motion has to be done numerically.
The procedure is summarized as follows and the same as what has been used in the literature to solve the coupled linear equations of motion for multi-field models.

We first set the $\delta\theta_\bk$ field to be in the Bunch-Davies (BD) vacuum and $\delta\sigma_\bk=0$. Solving the coupled differential equations leads to the value of $\delta\theta_\bk$ at the end of inflation, which we denote as $u_{\rm end}$ and will refer to as the ``$u$-contribution". Then, we set $\delta\theta_\bk=0$ and the $\delta\sigma_\bk$ field to be in the BD vacuum. This leads to another value of $\delta\theta_\bk$ at the end of inflation, which we denote as $v_{\rm end}$ and will refer to as the ``$v$-contribution". Because the massive field completely decays away after several efolds, the value of $\delta\theta_\bk$ at the end of inflation is all we need to evaluate the curvature perturbations. Using the time-delay formula between the curvature perturbation $\zeta$ and the field perturbation $\delta\theta$, $\zeta = -\delta\theta/\dot\theta$, the power spectrum is given by
\bea
P_\zeta = \frac{k_1^3}{2\pi^2} \frac{H^2}{\dot\theta_0^2}
( |u_{\rm end}|^2 + |v_{\rm end}|^2 ) ~,
\label{power_spectrum_uv}
\eea
where the background parameters $H$ and $\dot\theta_0$ are also evaluated at the end of inflation. As one can show, this procedure gives the complete tree-level non-perturbative results for the power spectrum.

In the following we separately investigate models that have smaller and larger values of $\epsilon$.
For the model examples we are most interested in in this paper,
these two cases roughly correspond to the small field inflation case and large field inflation case, respectively. The results for the two cases are qualitatively different.

\subsubsection{Small field case}
\label{Sec:small_field}

In the small $\epsilon$ limit, we can ignore all terms that are suppressed by $\epsilon$ in (\ref{L2_leading}). These terms can be most easily identified from the estimates in Table \ref{Table:Estimate} in Appendix \ref{App:quadratic_action}. In particular, we note that the coupling terms between $\delta\theta$ and $\delta\sigma$ are all negligible, so the two fields are decoupled in this limit.\footnote{The other parameters that should be held fixed when sending $\epsilon_0\to 0$ can be explicitly seen in (\ref{two_couplings}).} In addition, the $\delta\theta^2$ terms coming from the gravity sector are also suppressed. Therefore, in this limit the features in the curvature perturbation $\delta\theta$ is only contributed by the oscillations in the background field through the direct coupling, and the massive field quantum fluctuations do not contribute.
The equations of motion are dramatically simplified to the following single equation,
\begin{align}
\ddot{\delta\theta} +
\left[ 3H + \frac{2\dot\sigma_0}{\tR + \sigma_0} \right]
\dot{\delta\theta}
+
\left[ \frac{k^2}{a^2} + \frac{V''_{\rm sr}}{(\tR + \sigma_0)^2} \right]
\delta\theta =0 ~.
\label{EOM_smallfield}
\end{align}
Of course only the $u$-contribution is left in this case.

\begin{figure}[t]
  \centering
  \includegraphics[width=0.8\textwidth]{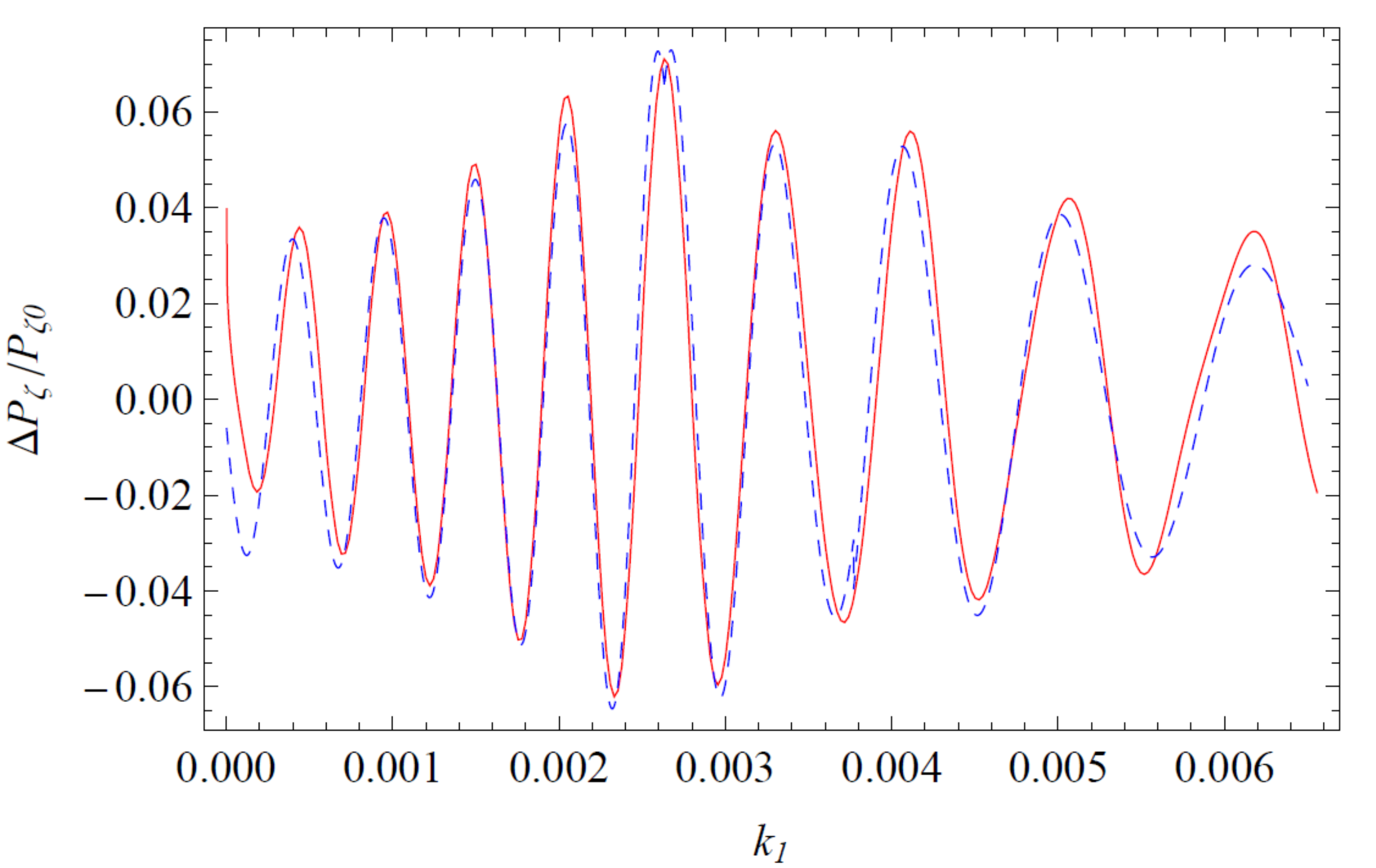}
  \caption{\label{Fig:smallfield_full} \small The numerical result of a full Standard Clock signal using the example in Fig.~\ref{Fig:bkgd_figs}. The solid line is the numerical result. The dashed line is the template (\ref{Template_twoparameter}) with $k_r=0.0055$ and $C=0.065$.
  The sharp feature starts from $k_1=k_0/2=8.7\times 10^{-5}$, the clock signal starts from $k_1=k_r/2=2.7\times 10^{-3}$.
  Note that the values of $k_1$ in this and later similar figures only have relative meanings. They will be rescaled in the data fitting.
  }
\end{figure}

An example is presented in Fig.~\ref{Fig:smallfield_full} in terms of the fractional correction $\Delta P_\zeta/P_{\zeta 0}$, where $P_{\zeta 0}$ is the featureless power spectrum. In CMB, this correction manifests as the residuals of the angular temperature and polarization power spectra after the $\Lambda$CDM model is subtracted.
As we can see, the sinusoidal sharp feature signal, starting from $k_1 \sim 8.7\times 10^{-5}$, smoothly connects with the clock signal, starting from $k_1\sim 2.7\times 10^{-3}$.
The overall pattern shown in Fig.~\ref{Fig:smallfield_full} is quite robust against change of parameters, as determined by the general properties in Sec.~\ref{Sec:main_properties}.
In this example, the clock oscillates for about 5 times in a Hubble time $H^{-1}$, while the number of oscillations of the clock signal shown in Fig.~\ref{Fig:smallfield_full} is about 4, so all signals show in Fig.~\ref{Fig:smallfield_full} are generated within a Hubble time. Generally, within a Hubble time, the number of oscillations generated in the clock signal is $\omega/(2\pi H) \gg 1$, which is enough to pin down the range of $p$.

Ideally, to compare with data in MCMC, we would like to have an analytical expression for these full signals which can accommodate the changes of all parameters. Practically, due to the smooth transition between the two completely different running behaviors, we have not figured out a way to achieve this goal. On the other hand, directly varying model parameters and feeding MCMC with the numerical results may be too computationally expensive. As a compromise, the best we can do so far is to fix certain parameters according to the best-fit results in the clock-signal-only MCMC analysis in Sec.~\ref{Sec:MCMC_clock}; then we mimic the numerical results using analytical templates which only allow the variation of the rest of the parameters.
The parameter that is most difficult to accommodate analytically is the clock frequency $\Omega$. So in different numerical simulations, including Fig.~\ref{Fig:smallfield_full}, we fix certain model parameters to fix $m_\sigma$, so that $\Omega \approx 30$ which is approximately the best-fit value from Sec.~\ref{Sec:MCMC_clock}.
Once this is fixed, the main remaining freedoms are the overall rescaling of the $k$-scale and the overall amplitude of the oscillatory feature, which we can describe by two parameters, $k_r$ and $C$, using the following template,
\bea
\frac{\Delta P_\zeta}{P_{\zeta0}} =
\left\{
\begin{array}{ll}
\high{
C \left[ 7\times 10^{-4} \left( \frac{2k_1}{k_0} \right)^2 + 0.5 \right]
\cos \left[ \frac{2k_1}{k_0} + 0.55\pi \right] ~,
}
&
k_1 < k_a ~,
\\
\high{
\frac{14}{13} C \left( \frac{2k_1}{k_r} \right)^{-3/2}
\sin \left[ \Omega \ln \frac{2k_1}{k_r} + 0.75\pi \right] ~,
}
&
k_b >k_1 \ge k_a ~,
\\
\high{
\frac{19}{13} C \left( \frac{2k_1}{k_r} \right)^{-3/2}
\sin \left[ \Omega \ln \frac{2k_1}{k_r} + 0.75\pi \right] ~,
}
&
k_1 \ge k_b ~,
\end{array}
\right.
\label{Template_twoparameter}
\eea
where
\bea
k_0 = \frac{k_r}{1.05\Omega} ~, ~~~
k_a = \frac{67}{140}k_r ~, ~~~
k_b = \frac{24}{35}k_r ~, ~~~
\Omega=30 ~.
\eea
This template is referred to as T2 in Sec.~\ref{Sec:MCMC_full}.
The comparison between the numerical result and the template is also demonstrated in Fig.~\ref{Fig:smallfield_full}. In the scales of interest, the difference is negligible. The smooth transition between the sharp feature signal and the clock signal are manifest analytically in this template.
The general function which decides the clock pattern in (\ref{clock_signal}) reduces to $\ln(2k_1/k_r)$ here because we have taken the exponential inflationary background.

\medskip

We further elaborate our understanding of this model by investigating the following issues.
\begin{itemize}

\item
{\em Analytical estimate of the amplitude}

In the simple limit (\ref{EOM_smallfield}), we can compute the clock signal analytically. The Lagrangian (\ref{L2_leading}) in this case becomes
\bea
\CL_2 \approx \frac{a^3}{2} (\tR+\sigma_0)^2
\left[ \dot{\delta\theta}^2 - \frac{1}{a^2} (\partial_i\delta\theta)^2 \right] ~,
\label{small_field_L2}
\eea
where we have further ignored the small mass term.
In the inflationary background, the evolution of the clock field $\sigma$ can be solved from Eq.~(\ref{massive_eom}),
\bea
\sigma_0 = \sigma_A e^{-\frac{3}{2} H (t-t_0)} \sin (m_\sigma t+{\rm phase}) ~, ~~~ t\ge t_0 ~.
\eea
Because the oscillation amplitude $\sigma_A$ is much smaller than the radius of the turning trajectory $\tR$, we can treat the $\CO(\sigma)$ terms in (\ref{small_field_L2}) as perturbations and compute their correction to the leading order power spectrum using the in-in formalism. This is the same type of computation as in Ref.~\cite{Chen:2011zf}.
The leading and subleading Hamiltonians are given by
\begin{align}
\CH_0 &= \frac{a^3}{2} \tR^2 \left[ \dot{\delta\theta}^2 + \frac{1}{a^2} (\partial_i \delta\theta)^2 \right] ~,
\nonumber \\
\CH_2^I &= -a^3 \tR \sigma_0 \left[ \dot{\delta\theta}^2 - \frac{1}{a^2} (\partial_i \delta\theta)^2 \right] ~.
\end{align}
Treating $\CH_2^I$ as perturbations in the in-in formalism, we get
\begin{align}
\frac{\Delta P_\zeta}{P_{\zeta0}} =& -2 i \int_{-\infty}^0 d\tau a^2 \tR \sigma_0
\left({u_{k_1}'}^2 - k_1^2 u_{k_1}^2 \right) + {\rm c.c.}
\nonumber \\
=&
\sqrt{2\pi} \frac{\sigma_A}{\tR}
\left(\frac{m_\sigma}{H}\right)^{1/2}
\left( \frac{2k_1}{k_r} \right)^{-3/2}
\sin \left( \frac{m_\sigma}{H} \ln \frac{2k_1}{k_r} + {\rm phase} \right) ~,
\label{small_field_clock_analytical}
\end{align}
where $k_1$ and $k_r$ are the same as those defined below Eq.~(\ref{clock_signal}). The $u_k$ is the mode function for the $\delta\theta$ field. In this calculation, the only behavior of $u_k$ needed is its subhorizon BD behavior given in (\ref{BD_vacuum}). The analytical result (\ref{small_field_clock_analytical}) takes the same form as the last clock signal component in the template (\ref{Template_twoparameter}). To compare their amplitudes, we use the example in Fig.~\ref{Fig:bkgd_figs} and \ref{Fig:smallfield_full} and compute the initial amplitude of the clock signal analytically by estimating $\sigma_A \approx \sigma_f$, $\tR \approx 2.05$, $m_\sigma/H \approx 30$. We get
$\Delta P_\zeta/P_{\zeta0} \approx 0.11$ at $k_1=k_r/2$. This matches well with the clock component in the template (\ref{Template_twoparameter}) -- evaluating the last line at  $k_1=k_r/2$ gives $\Delta P_\zeta/P_{\zeta0} \approx 0.095$.

\item
{\em Effect of the plateau shape $m_0$.}

We have used (\ref{V_sigma}) to model the potential dip for the $\sigma$-field. As an initial condition, we placed the inflaton at the plateau of (\ref{V_sigma}) near the potential dip so that the inflaton can drop into the dip after a few efolds of rolling. We have also provided the motivation for such model-building. Clearly, many different shapes of potential may be used to model such a tachyonic falling process, and this is only one simple example. Certain aspects of this example may not fully represent the situation in the more realistic model-building. It is an interesting question how these variations may alter the theoretical predictions.

For example, the shape of the plateau in (\ref{V_sigma}) is very flat faraway from the dip. Realistically, potentials in inflationary background receive corrections that are of the form $\sim H^2 \sigma^2$. To test the effect of this term, we replace (\ref{V_sigma}) with
\bea
V_\sigma = V_{\sigma 0} \left[ 1- \exp(-\sigma^2/\sigma_f^2) \right]
+ \half m_0^2 \sigma^2 ~,
\label{V_sigma_2}
\eea
where $m_0 \sim \CO(H)$.

\begin{figure}[p]
  \centering
  \includegraphics[width=0.49\textwidth]{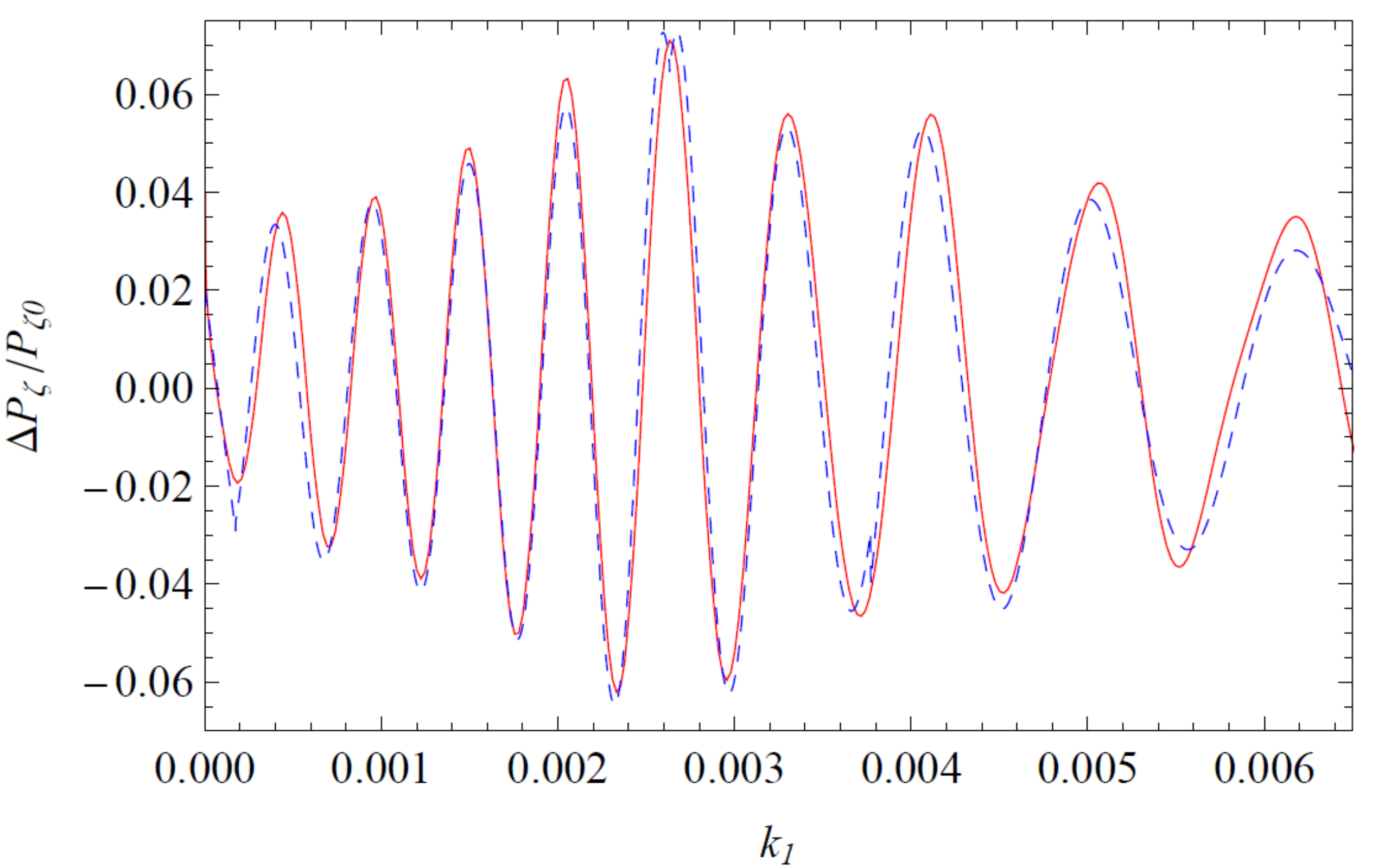}
  \includegraphics[width=0.49\textwidth]{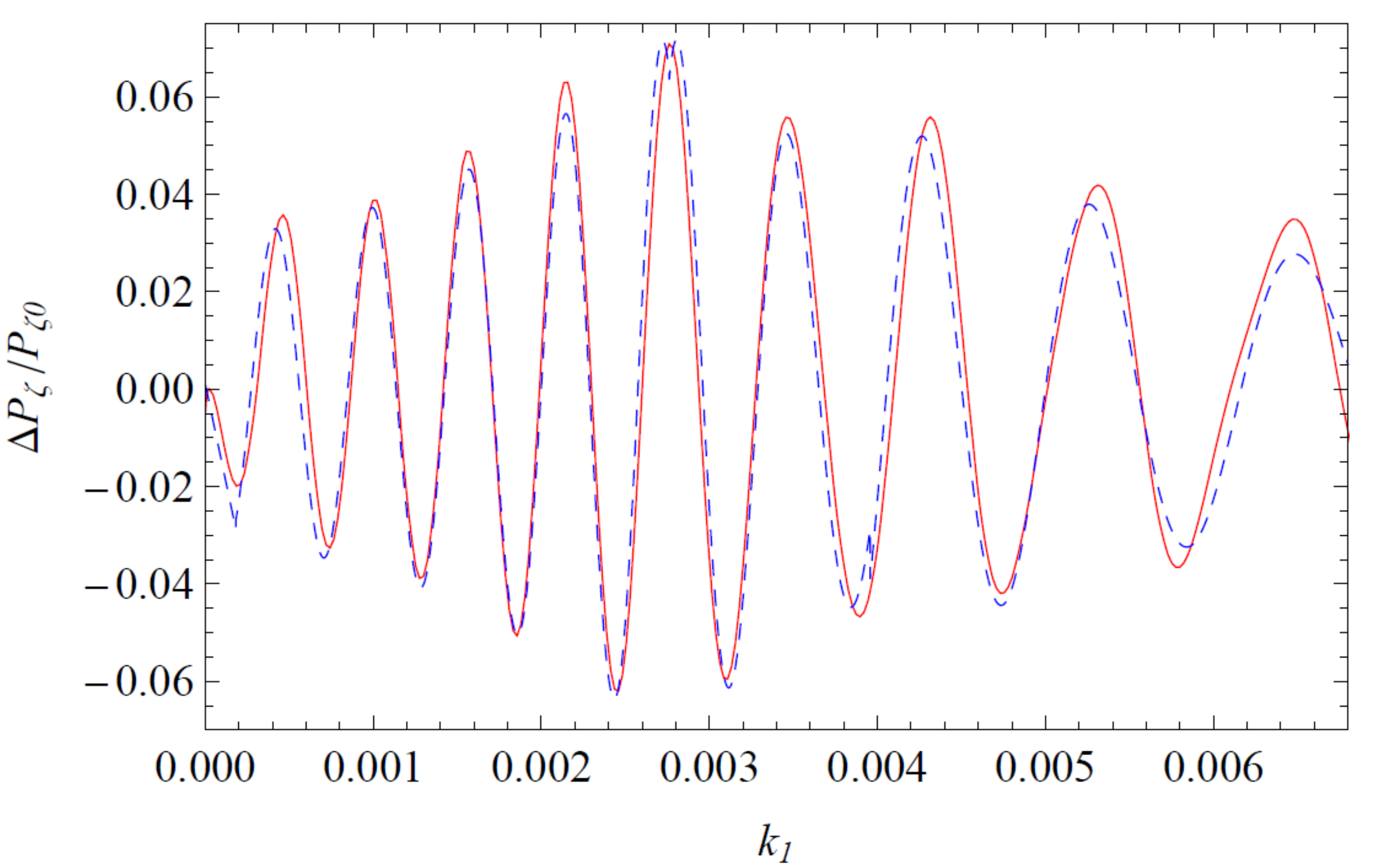}
  \includegraphics[width=0.49\textwidth]{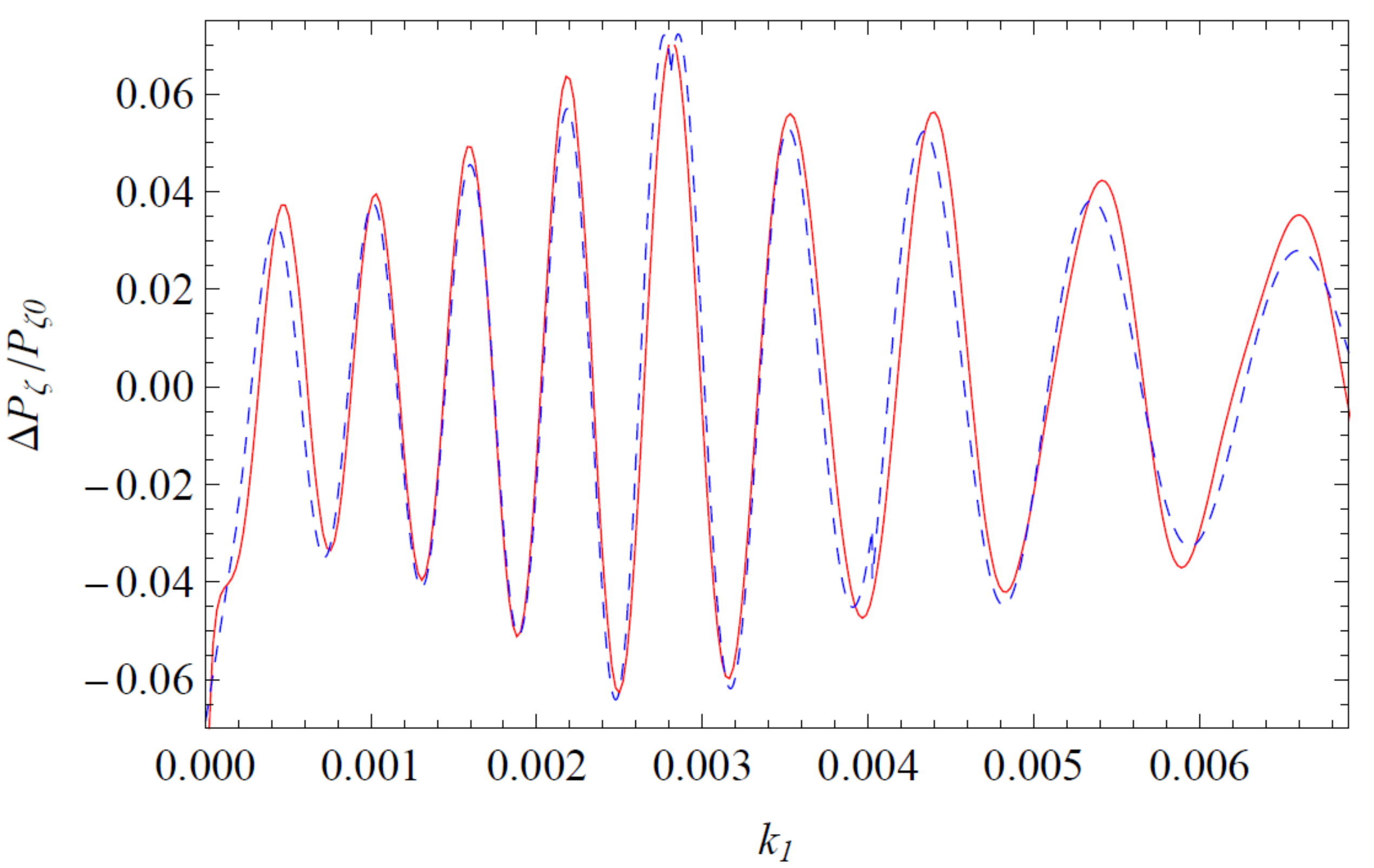}
  \includegraphics[width=0.49\textwidth]{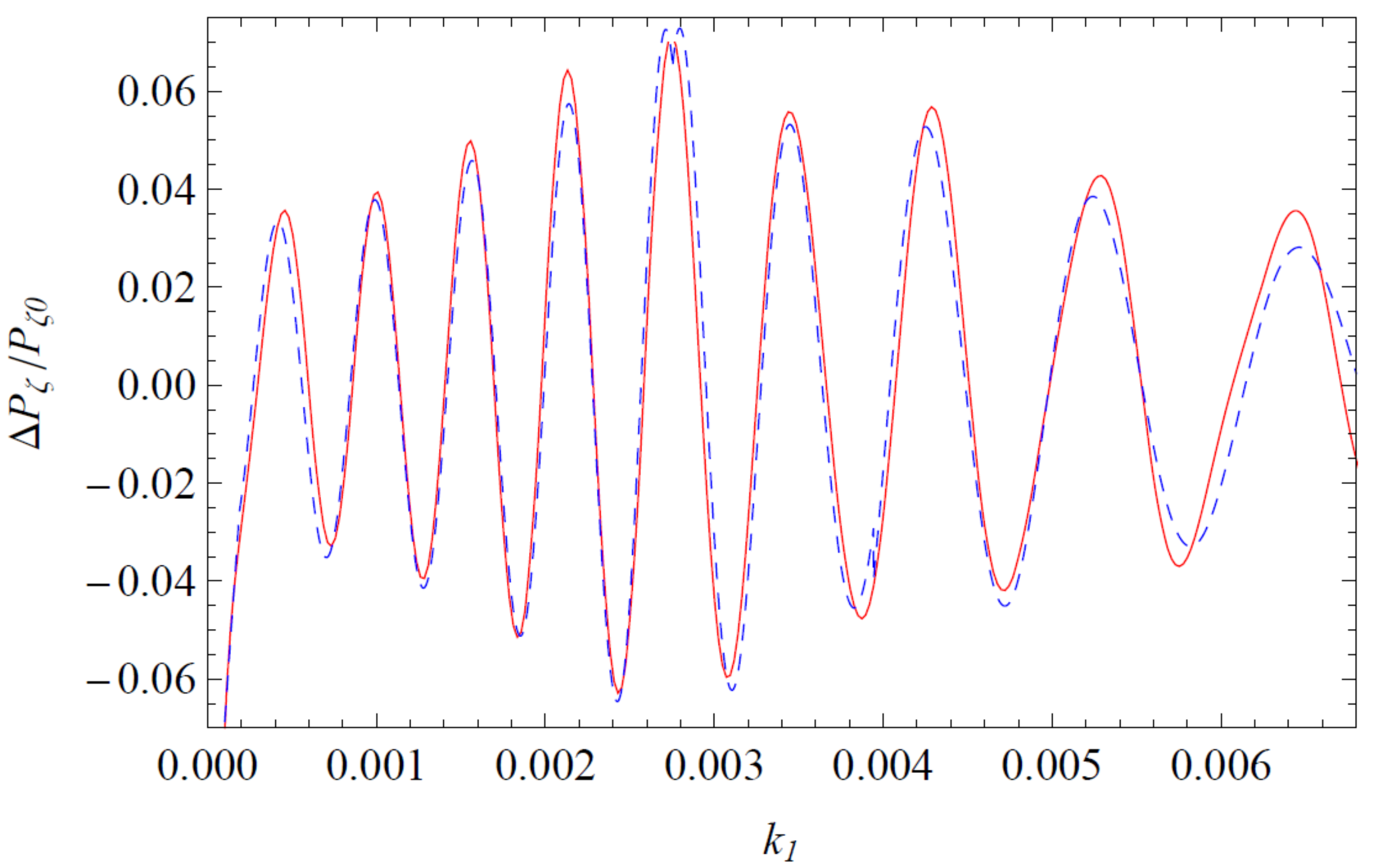}
  \caption{\label{Fig:effect_m0} \small The effect of the plateau shape in the $\sigma$-field potential (\ref{V_sigma_2}), parameterized by $m_0$, on the density perturbations in the small field case ($\epsilon_0 \approx 10^{-6}$).
  The solid lines are numerical results. The dashed lines are the template (\ref{template3}).
  The values of $m_0$ are:
  {\em upper-left}: $m_0=0$;
  {\em upper-right}: $m_0=0.4 H_0$;
  {\em lower-left}: $m_0=1.0 H_0$;
  {\em lower-right}: $m_0=1.5 H_0$.
  \\
  The other model parameters are:
  \\
  {\em upper-left panel}: $V_{\rm inf}=5.33\times 10^{-13}$, $V_{\sigma 0}=2.66\times 10^{-14}$, $\sigma_f = 1.64\times 10^{-2}$, $\tR=2.05$, $\beta=1.55\times 10^{-15}$;
  \\
  {\em upper-right panel}: $V_{\rm inf}=5.33\times 10^{-13}$, $V_{\sigma 0}=2.66\times 10^{-14}$, $\sigma_f = 1.64\times 10^{-2}$, $\tR=2.05$, $\beta=1.55\times 10^{-15}$;
  \\
  {\em lower-left panel}: $V_{\rm inf}=5.33\times 10^{-13}$, $V_{\sigma 0}=1.06\times 10^{-14}$, $\sigma_f = 1.04\times 10^{-2}$, $\tR=1.30$, $\beta=9.79\times 10^{-16}$;
  \\
  {\em lower-right panel}: $V_{\rm inf}=5.33\times 10^{-13}$, $V_{\sigma 0}=1.06\times 10^{-14}$, $\sigma_f = 1.04\times 10^{-2}$, $\tR=1.30$, $\beta=9.79\times 10^{-16}$.
  \\
  We have chosen the initial conditions so that the number of efolds of the first inflationary stage is approximately six; also, the parameters are chosen to keep $\Delta P_\zeta/P_{\zeta0}$ the same, for better visual comparison.
  }
\end{figure}

A series of examples with increasing $m_0^2$ are shown in Fig.~\ref{Fig:effect_m0}. As we can see, this term does not have significant effect on the clock signal. The main change is in the initial part of the sharp feature signal in the largest scales. The main effect on these scales are an overall suppression. This suppression is stronger for larger $m_0^2$. This modification can be understood because this term modifies the initial part of the non-attractor phase by increasing the initial rolling velocity of the inflaton. This larger initial velocity suppresses the density perturbations at the largest scales. To compare this modification with data, we can modify the template (\ref{Template_twoparameter}) by adding another parameter $b$ to describe the large scale suppression in the following template,
\bea
\frac{\Delta P_\zeta}{P_{\zeta0}} =
\left\{
\begin{array}{ll}
\high{
-0.45 C - b (1- \frac{k_1}{k_0}) ~,
}
&
k_1<k_0 ~,
\\
\high{
C \left[ 7\times 10^{-4} \left( \frac{2k_1}{k_0} \right)^2 + 0.5 \right]
\cos \left[ \frac{2k_1}{k_0} + 0.55\pi \right] ~,
}
&
k_a >k_1 \ge k_0 ~,
\\
\high{
\frac{14}{13} C \left( \frac{2k_1}{k_r} \right)^{-3/2}
\sin \left[ \Omega \ln \frac{2k_1}{k_r} + 0.75\pi \right] ~,
}
&
k_b >k_1\ge k_a ~,
\\
\high{
\frac{19}{13} C \left( \frac{2k_1}{k_r} \right)^{-3/2}
\sin \left[ \Omega \ln \frac{2k_1}{k_r} + 0.75\pi \right] ~,
}
&
k_1\ge k_b ~,
\end{array}
\right.
\label{template3}
\eea
where
\bea
k_0 = \frac{k_r}{1.05\Omega} ~, ~~~
k_a = \frac{67}{140}k_r ~, ~~~
k_b = \frac{24}{35}k_r ~, ~~~
\Omega=30 ~.
\eea
This template is referred to as T3 in Sec.~\ref{Sec:MCMC_full}.

\item
{\em Effect of the magnitude of $\epsilon_0$.}

Let us approximate and denote the slow-roll parameter $\epsilon$ in the attractor phase as
\bea
\epsilon_0 = \frac{\dot\theta_0^2 \tR^2}{2 \mpl^2 H^2} ~.
\eea
So far we have computed the power spectrum in the small $\epsilon_0$ limit. In this limit, the fields $\delta\theta$ and $\delta\sigma$ are decoupled and the other perturbative terms from the gravity sector are also negligible. As we increase $\epsilon_0$ towards a certain critical value, these terms start to become important. In particular, we can no longer ignore the coupling terms, and the quantum fluctuation $\delta\sigma$ starts to contribute to the curvature perturbation. We then have to solve the two coupled equations of motion following the procedure prescribed above.

\begin{figure}[p]
  \centering
  \includegraphics[width=0.49\textwidth]{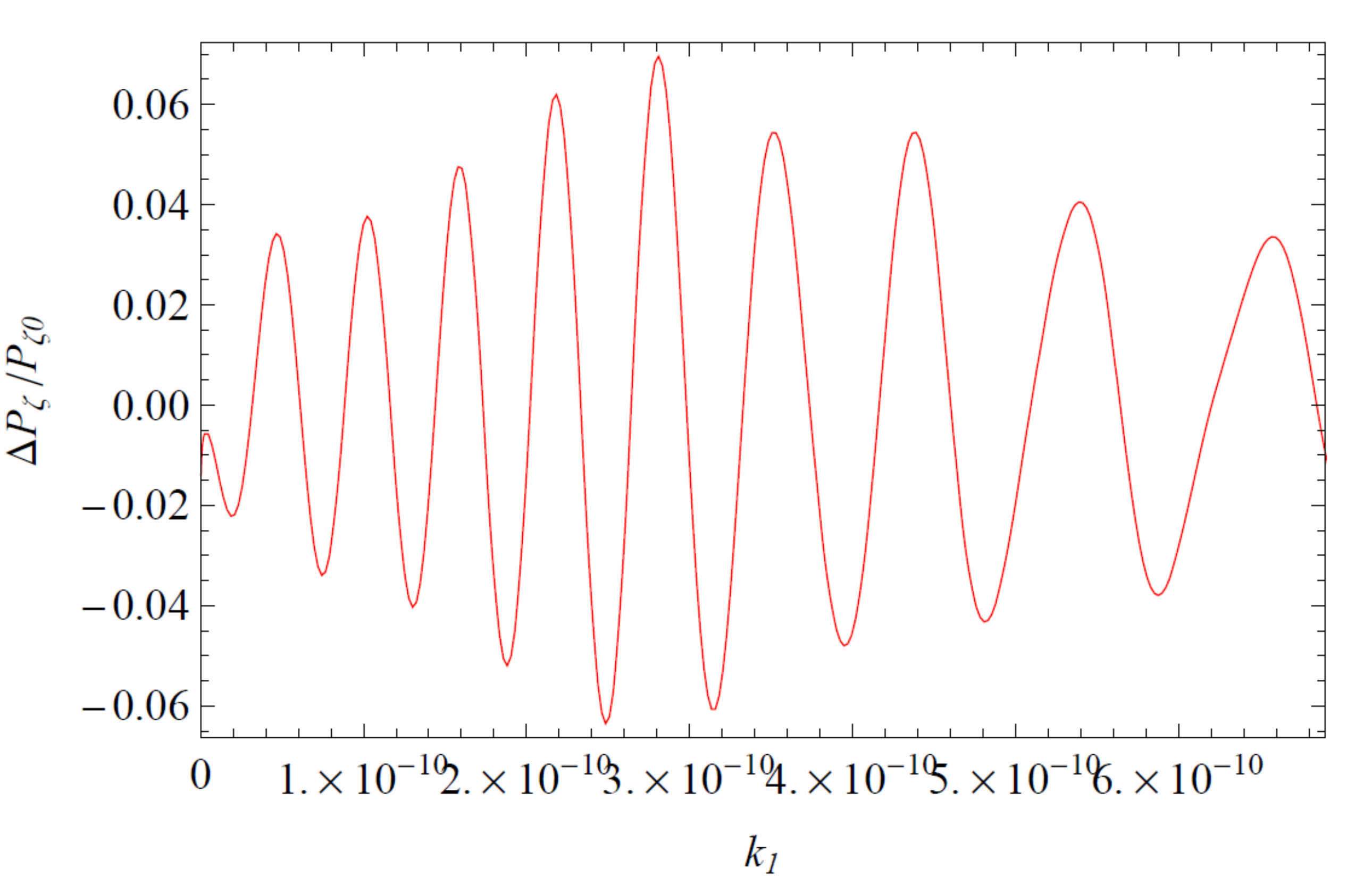}
  \includegraphics[width=0.49\textwidth]{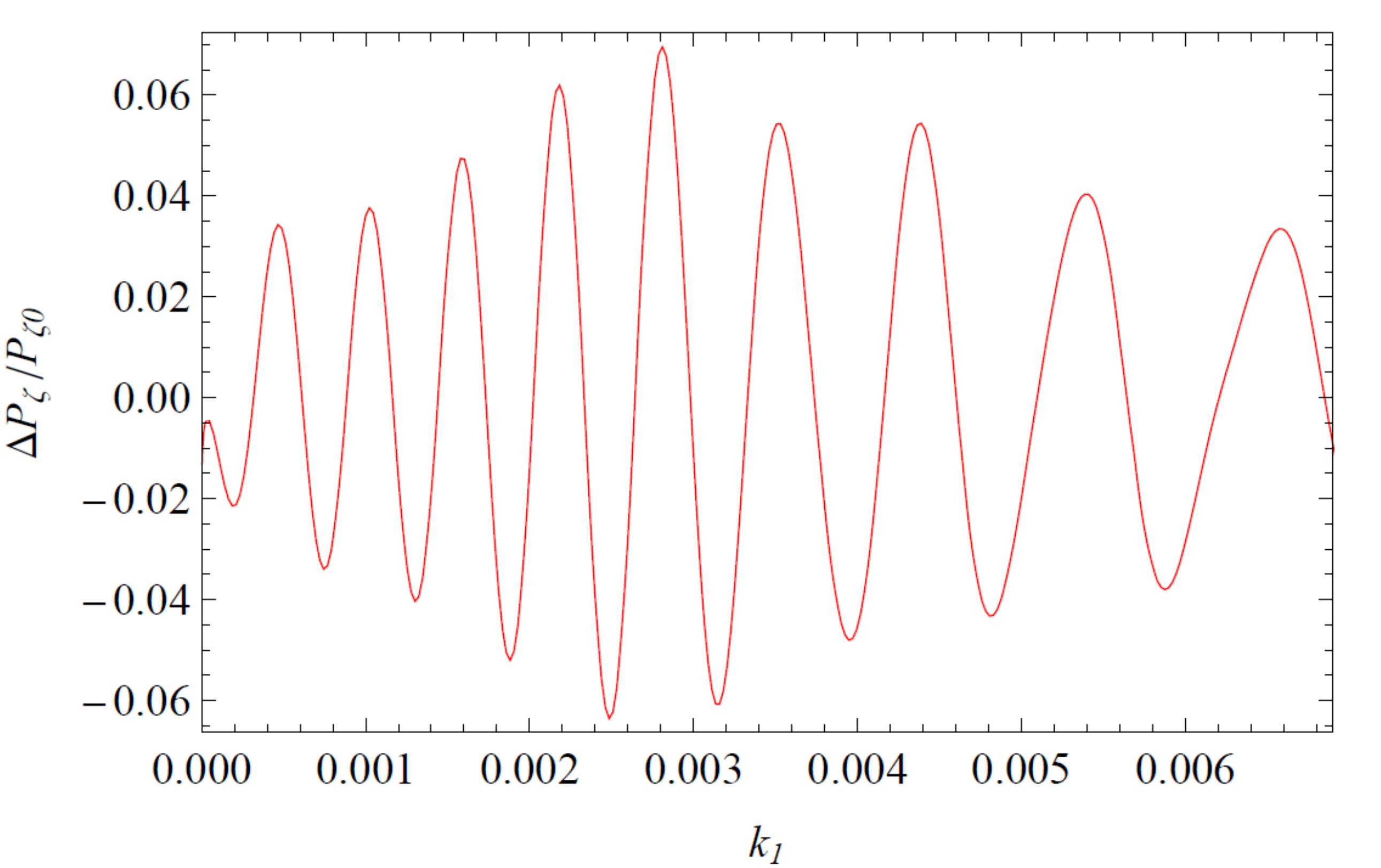}
  \includegraphics[width=0.49\textwidth]{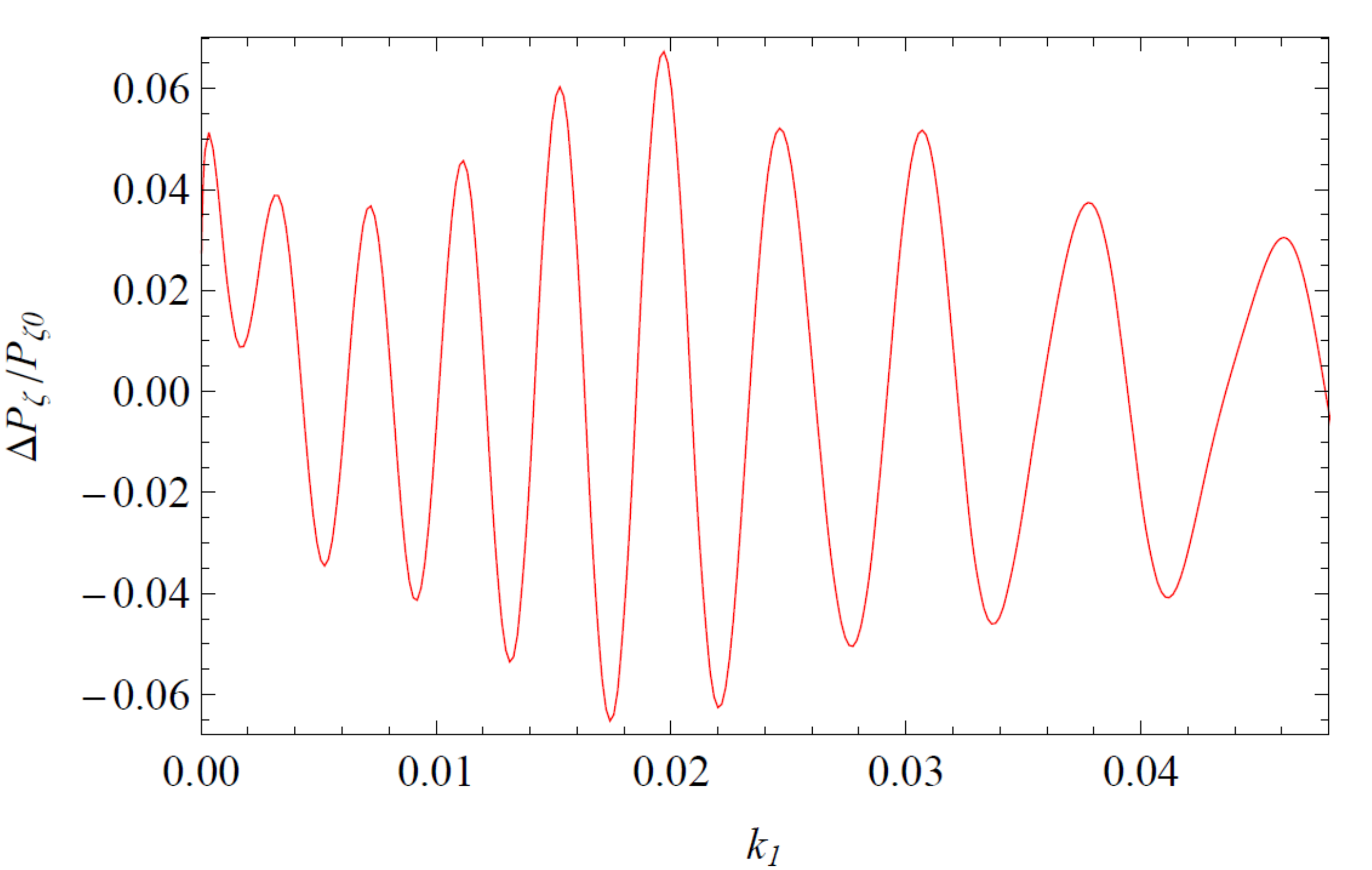}
  \includegraphics[width=0.49\textwidth]{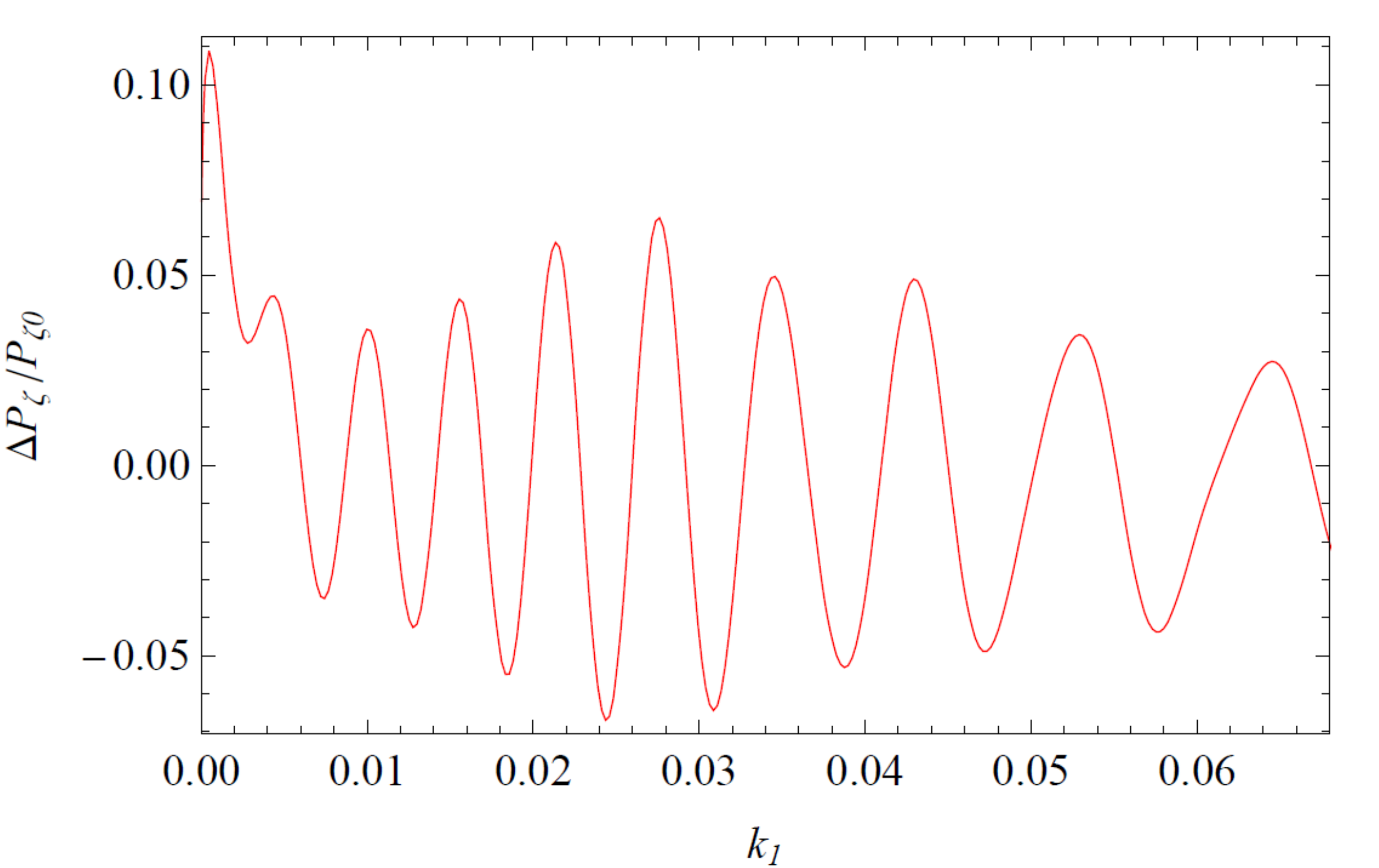}
  \caption{\label{Fig:effect_epsilon} \small The effect of increasing $\epsilon_0$ on the density perturbations. The attractor values of $\epsilon$ are:
  {\em upper-left}: $\epsilon_0 = 10^{-20}$;
  {\em upper-right}: $\epsilon_0 = 10^{-6}$;
  {\em lower-left}: $\epsilon_0 = 5\times 10^{-5}$;
  {\em lower-right}: $\epsilon_0 = 10^{-4}$.
  \\
  The model parameters are:
  \\
  {\em upper-left panel}: $V_{\rm inf}=5.33\times 10^{-27}$, $V_{\sigma 0}=2.66\times 10^{-28}$, $\sigma_f = 1.64\times 10^{-2}$, $\tR=2.05$, $\beta=1.55\times 10^{-36}$, $m_0=0.5 H_0$.
  \\
  {\em upper-right panel}: $V_{\rm inf}=5.33\times 10^{-13}$, $V_{\sigma 0}=2.66\times 10^{-14}$, $\sigma_f = 1.64\times 10^{-2}$, $\tR=2.05$, $\beta=1.55\times 10^{-15}$, $m_0=0.5 H_0$.
  \\
  {\em lower-left panel}: $V_{\rm inf}=2.66\times 10^{-11}$, $V_{\sigma 0}=1.33\times 10^{-12}$, $\sigma_f = 1.64\times 10^{-2}$, $\tR=2.05$, $\beta=5.47\times 10^{-13}$, $m_0=0.5 H_0$.
  \\
  {\em lower-right panel}: $V_{\rm inf}=5.33\times 10^{-11}$, $V_{\sigma 0}=2.66\times 10^{-12}$, $\sigma_f = 1.64\times 10^{-2}$, $\tR=2.05$, $\beta=1.55\times 10^{-12}$, $m_0=0.5 H_0$.
  }
\end{figure}

A series of examples are shown in Fig.~\ref{Fig:effect_epsilon} to demonstrate the effect of growing $\epsilon_0$. As we can see, in these examples, as $\epsilon_0 < 10^{-4}$, the change of $\epsilon_0$ has negligible effect on the full Standard Clock signal. For $\epsilon_0 \gtrsim 10^{-4}$, the amplitude of the sharp feature signal starts to grow and becomes larger than the amplitude of the clock signal. As we will see in the next subsection, this trend continues in the large field models.

The examples we showed here all have $\omega/H \approx 30$ because this is the oscillation frequency of the best-fit model in Sec.~\ref{Sec:MCMC_clock}. With this parameter fixed, there are still several other parameters that can be varied. So one may wonder, for a given $\epsilon_0$ and a fixed amplitude of the clock signal, whether there exists some parameter space in which the coupling terms can still be made decoupled parametrically. Since there are two leading coupling terms in the Lagrangian \eqref{L2_leading} according to the Table~\ref{Table:Estimate} in Appendix~\ref{App:quadratic_action}, let us ask if we can tune model parameters so that both terms vanish while fixing $\epsilon_0$ and the amplitude of the clock signal.

In Table~\ref{Table:Estimate} the strength of the two terms are estimated as $\sqrt{\epsilon_0}(m_\sigma/H)(\mpl/\tR)$ and
$\sqrt{\epsilon_0}(m_\sigma/H)^2(\sigma_A/\mpl)$, respectively, where $\sigma_A$ is the initial oscillation amplitude of the clock field $\sigma$.
To relate variable parameters to fixed quantities, we note the following relations:
\begin{align}
& V_{\sigma 0} \sim m_\sigma^2 \sigma_A^2 \sim \dot\sigma_A^2 ~,
\nonumber \\
& V_{\rm inf} \sim H^2 \mpl^2 ~,
\end{align}
where $V_{\sigma 0}$ is the depth of the potential dip in the $\sigma$-field potential (\ref{V_sigma}) and (\ref{V_sigma_2}), $V_{\rm inf}$ is the inflationary energy, and we require $V_{\sigma 0} \ll V_{\rm inf}$.
Use these relations, we can write
\begin{align}
& \sqrt{\epsilon_0} \frac{m_\sigma}{H} \frac{\mpl}{\tR}
\sim \sqrt{\epsilon_0} \frac{(m_\sigma/H)^2(\sigma_A/\tR)}{\sqrt{V_{\sigma 0}/V_{\rm inf}}} ~,
\nonumber \\
& \sqrt{\epsilon_0} \left(\frac{m_\sigma}{H}\right)^2 \frac{\sigma_A}{\mpl} \sim
\sqrt{\epsilon_0} \left(\frac{m_\sigma}{H}\right) \sqrt{\frac{V_{\sigma 0}}{V_{\rm inf}}} ~.
\label{two_couplings}
\end{align}
Because, according to (\ref{small_field_clock_analytical}), the amplitude of the clock signal is approximately given by
\bea
\frac{\Delta P_\zeta}{P_{\zeta0}} \sim \sqrt{2\pi}
\frac{\sigma_A}{\tR} \sqrt{\frac{m_\sigma}{H}} ~,
\label{Clock_amplitude_estimate}
\eea
fixing the clock signal amplitude $\Delta P_\zeta/P_{\zeta0} \sim 10^{-1} - 10^{-2}$ means that we fix
$\sigma_A/\tR \sim 10^{-2}-10^{-3}$. Now we can see that it is impossible to make both coupling terms in (\ref{two_couplings}) small parametrically, once we fix $\epsilon_0$, $m_\sigma/H$ and $\Delta P_\zeta/P_{\zeta0}$. The minimum coupling value is obtained when the two couplings are equal, namely, when
\bea
\frac{V_{\sigma 0}}{V_{\rm inf}} \sim \frac{m_\sigma}{H} \frac{\sigma_A}{\tR} \sim \CO(0.1-0.01) ~.
\label{minimum_coupling_point}
\eea
This is roughly the relative size of the $\sigma$-field potential dip we took in the numerical simulation.
We can also use values other than (\ref{minimum_coupling_point}), in which case one of the two coupling terms is more important following (\ref{two_couplings}).
From the above analyses, we can see that, as $\epsilon_0$ increases, to keep the same observable clock signal, the coupling terms have to become increasingly important. For the best-fit example $m_\sigma/H\approx 30$, the size of the minimum coupling terms is of order
$\sim (m_\sigma/H)^{3/2} (\sigma_A/\tR)^{1/2} \sqrt{\epsilon_0} (a^3 H^2 \tR \delta\theta \delta\sigma)
\sim \CO(10) \sqrt{\epsilon_0} (a^3 H^2 \tR \delta\theta \delta\sigma)$,
where we have used (\ref{two_couplings}) and (\ref{minimum_coupling_point}).
From the numerical simulation, we observe that this coupling term becomes important when $\epsilon_0 \gtrsim 10^{-4}$. Coincidentally this critical $\epsilon_0$ value is also roughly the dividing line between the small field and large field inflation models.
However, note that this is a coincidence because the critical value of $\epsilon_0$ sensitively depends on the value of $\omega/H$, which we take as $\sim 30$ here. If we take a different $\omega$ value, the coupling terms, along with their associated observational effects we discuss below, can become important even for small field inflation models.

\item
{\em Relative amplitude between the sharp feature signal and clock signal.}

We note in the previous examples that the coupling terms, which become important when $\epsilon_0$ exceeds a critical value, have stronger effect on the sharp feature signal. The amplitude of the sharp feature signal grows much more quickly than that of the clock signal.
We can understand this observation by reminding the resonance mechanism in the generation process of the clock signal and separately considering the $u$ and $v$-contribution defined above Eq.~(\ref{power_spectrum_uv}).

The resonance mechanism \cite{Chen:2008wn}, which plays an important role in amplifying the amplitude of the clock signal, relies on the coincidence of the frequencies between the background oscillation and quantum fluctuations. For a nearly massless field, such as $\delta\theta$, its frequency starts from a large value and evolves to nearly zero after horizon crossing, hence each mode which is deep enough within the horizon at the time of sharp feature has a chance to resonate with the oscillatory background.
In contrast, the frequency of the massive quantum fluctuation $\delta\sigma$ never becomes smaller than the background frequency. So the quantum fluctuation $\delta\sigma$ does not resonate with the background.

In the $u$-contribution, the amplitude of the clock signal is mainly contributed by the resonance between the background and the inflaton quantum fluctuation $\delta\theta$. As $\epsilon_0$ grows above the critical value, the coupling terms also start to contribute to both the sharp feature and clock signals.  This is because the coefficients of these coupling terms contain the background oscillation which can resonant with $\delta\theta$.

In the $v$-contribution, the coupling terms mostly contribute to sharp feature signal, because $\delta\theta$ is set to zero initially and hence does not have the oscillatory behaviour to resonate with the background. The $\delta\theta$ field will later be induced by the massive field quantum fluctuation $\delta\sigma$, but as explained the behavior of $\delta\sigma$ does not resonate with the background. This induced $\delta\theta$ is the main source of the large spike near the sharp feature signal. As we can see from the numerical results, this contribution to sharp feature signal has a peculiar scale-dependence and even tends to spoil the original sharp feature sinusoidal running behavior.

\end{itemize}

\begin{figure}[t]
  \centering
  \includegraphics[width=0.49\textwidth]{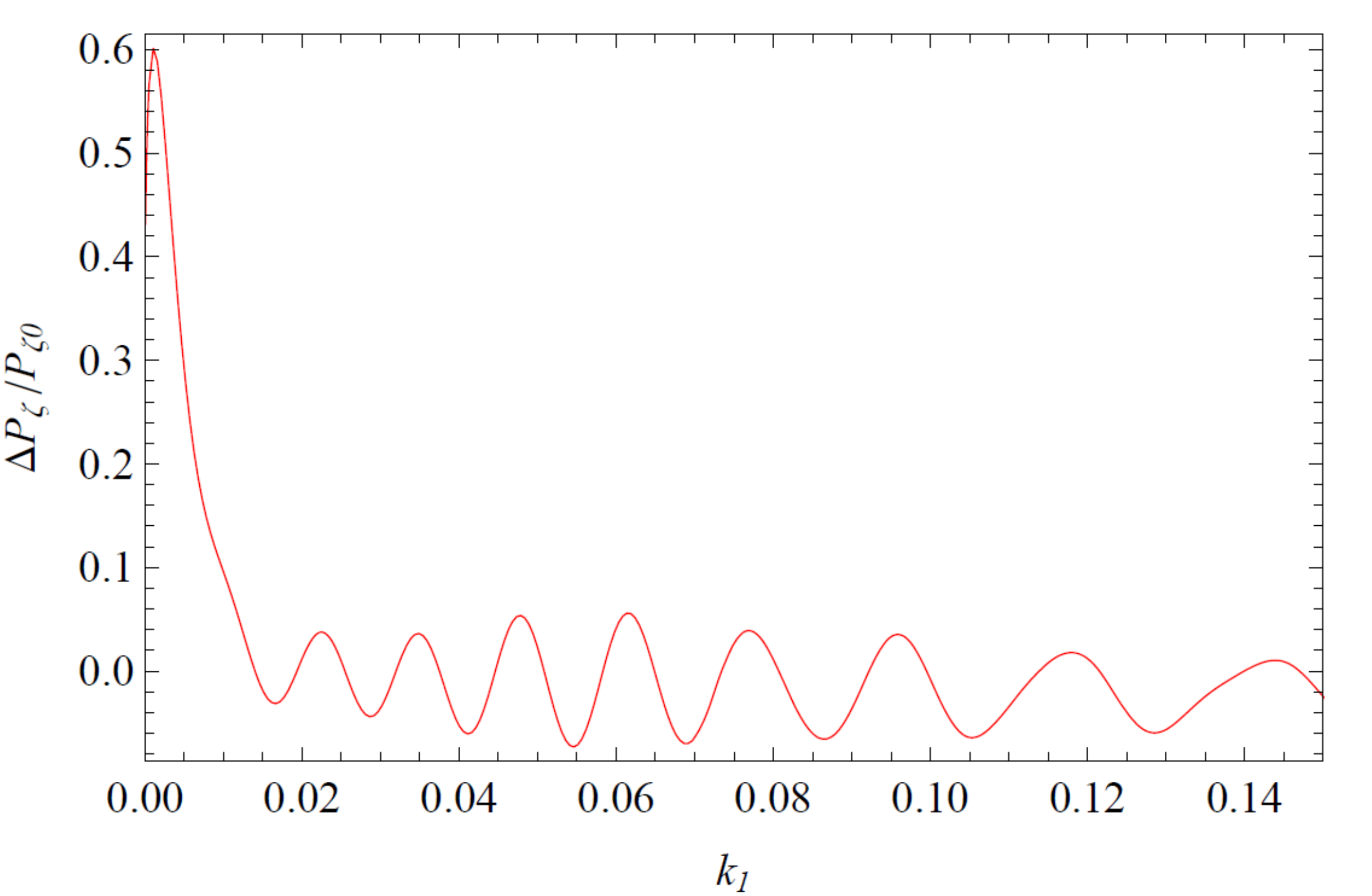}
  \includegraphics[width=0.49\textwidth]{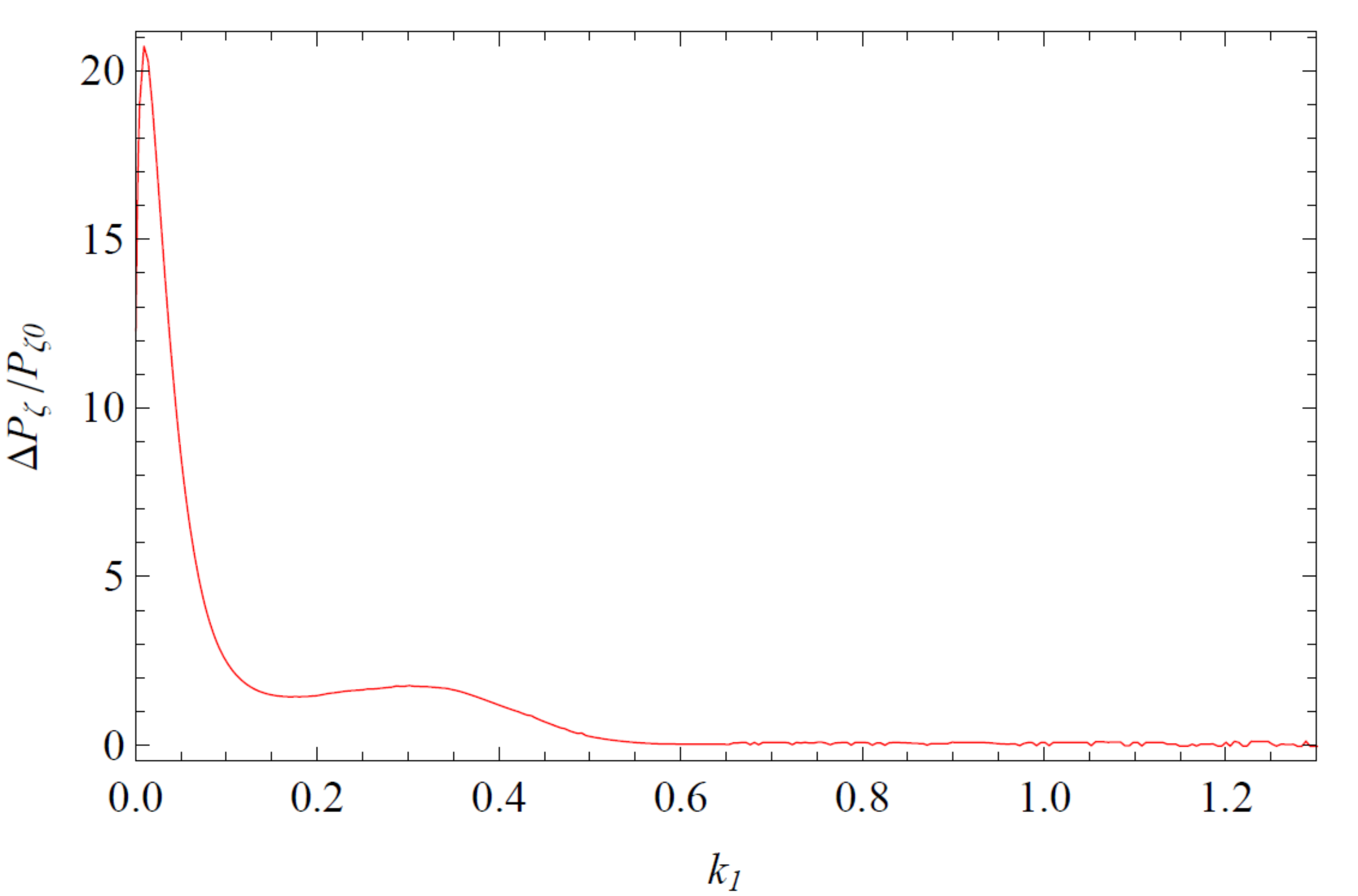}
  \caption{\label{Fig:Full_signal_large_field_epsilon} \small The effect of larger $\epsilon_0$ on the density perturbations. The attractor values of $\epsilon$ are: {\em left}: $\epsilon_0 = 5\times 10^{-4}$; {\em right}: $\epsilon_0 = 2.5 \times 10^{-2}$.
  \\
  The model parameters are:
  \\
  {\em left panel}: potential (\ref{V_smallfield}) is used; $V_{\rm inf}=2.66\times 10^{-10}$, $V_{\sigma 0}=1.33\times 10^{-11}$, $\sigma_f = 1.64\times 10^{-2}$, $\tR=2.05$, $\beta=1.73\times 10^{-11}$, $m_0=0.5 H_0$.
  \\
  {\em right panel}: potential (\ref{V_largefield}) is used; $m=1.01\times 10^{-5}$, $V_{\sigma 0}= 7.40\times 10^{-10}$, $\sigma_f = 2.32\times 10^{-2}$,
  $\tR = 23.2$, $m_0 = 0.5 H_0$.
  }
\end{figure}

\subsubsection{Large field case}
\label{Sec:Large_field}

To study the effect of even larger $\epsilon_0$, we continue to increase the value of $\epsilon_0$ in the potential (\ref{V_smallfield}). We also replace this potential with the following canonical large field potential
\bea
V_{\rm sr} = \half m^2 \tR^2 \theta^2 ~,
\label{V_largefield}
\eea
so that $\epsilon_0$ becomes of order $10^{-2}$. Two examples are shown in Fig.~\ref{Fig:Full_signal_large_field_epsilon}.
As we can see, the discussions in the previous subsection on the effect of the larger $\epsilon_0$ still apply here.
This is the case in which the coupling terms and the perturbative terms from the gravity sector become dominant. Ignoring them would lead to a qualitatively different result. The spike near the sharp feature signal becomes predominant as $\epsilon_0$ enters the large field region.
As discussed, this spike is contributed by the $v$-contribution and is due to the coupling between $\delta\theta$ and $\delta\sigma$. From the figures we can also see that its scale dependence is quite different from both the resonant running and the sinusoidal running.
Due to this spike, the amplitude of the clock signal becomes relatively negligible.
For this reason, in the next section where we apply the full signal to the data analyses, we concentrate on the small-field case with small $\epsilon_0$.

The nature of the sharp feature and the nature of the coupling between the clock field and the density-perturbation-source-field are the two most model-dependent aspects of the Standard Clock models. It is an interesting question whether we can construct large field inflation models with significant clock signals.

\section{MCMC analysis on the full signal}
\label{Sec:MCMC_full}
\setcounter{equation}{0}

In this section, we perform MCMC analysis on the full Standard Clock signal worked out in Sec.~\ref{Sec:Full_Model}.
The full signal includes both the sharp feature signal and the clock signal, and they are smoothly connected to each other. As mentioned, currently we do not have an analytical template which allows the variations of all parameters. What we can do so far is to fix some parameters in the model using the best-fit results of the clock-signal-only analysis in Sec.~\ref{Sec:MCMC_clock}. The parameter most difficult to vary in a full-model template is $\Omega$. To fix the value of $\Omega$, we tune the model parameters to fix the mass of the oscillating clock field in the numerical simulations. Once the $\Delta P_\zeta/P_{\zeta0}$ is produced numerically, the phase $\phi$ is also fixed.
We then propose a template that fits this numerical result. Since we constructed the full-model only for the case of exponential inflation, $p \gg 1$ is no longer a parameter.

Two templates are proposed in Sec.~\ref{Sec:Full_Model}.

The two-parameter template (T2) is given in (\ref{Template_twoparameter}). In this template, the variable parameters are the scale of the clock signal $k_r$ and the overall amplitude $C$.
Considering that we have effectively fixed two parameters in the simulation, this template effectively have 4 parameters. In the MCMC analyses, flat priors are given to $\log k_\mathrm{r}$ and $C$.

The three-parameter template (T3) is given in (\ref{template3}). This template has one additional parameter $b$ that controls the degree of large scale suppression as a consequence of changing the plateau shape of the potential.
Similarly, this template effectively has 5 parameters. In the MCMC analyses, only 3 are allowed to vary, and flat priors are given to $\log k_\mathrm{r}$, $C$ and $b$.

\begin{figure}[t]
  \centering
  \includegraphics[width=0.6\textwidth]{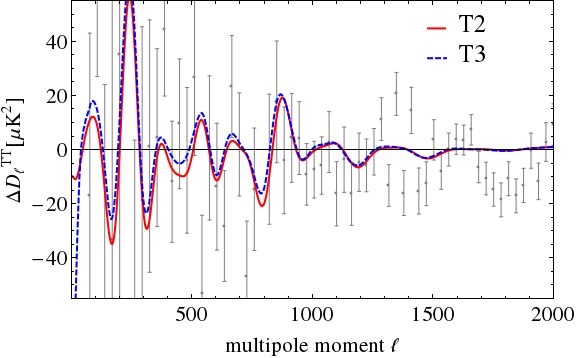}
  \caption{\label{fig:best-template} \small Corrections to the CMB temperature anisotropies from templates \eqref{Template_twoparameter} and \eqref{template3}. }
\end{figure}

The best-fit models for templates T2 and T3 are given in Table~\ref{tab:best-chi2-full}, and plotted in Fig.~\ref{fig:best-template}.
As expected, T2 and T3 are quite similar except at low $\ell<30$. T3 has a suppression which fits the observed deficit of power.
The most significant fit for both T2 and T3 is again the wiggle at around $\ell \sim 800$. Below $\ell<800$, both templates are essentially examples of the sinusoidal sharp feature. Above $\ell>800$, the inflationary clock signal starts to show up.

\begin{figure}[htbp]
  \centering
  \includegraphics[width=0.8\textwidth]{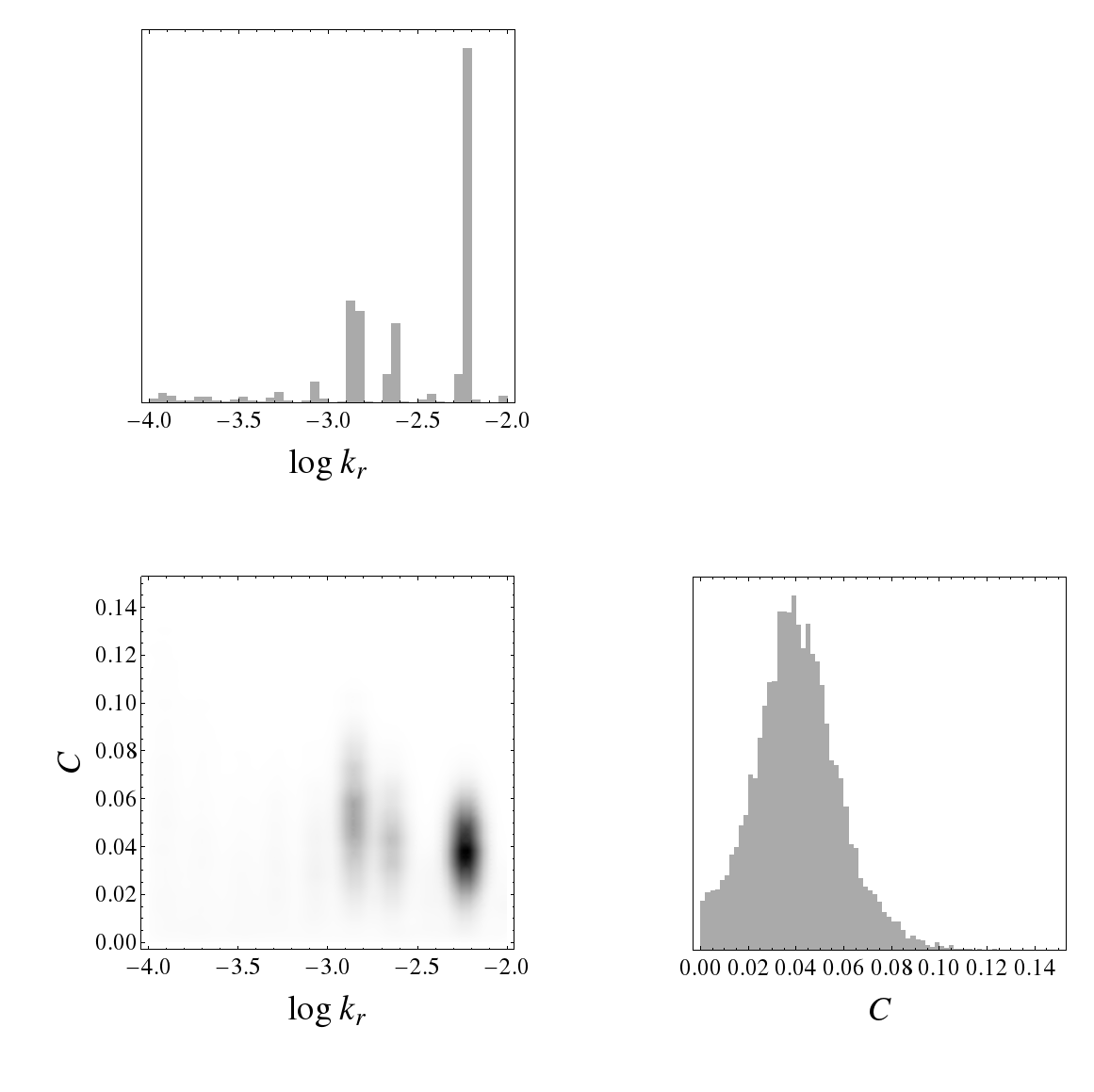}
  \caption{\label{fig:tp0} \small The triangle plot for template \eqref{Template_twoparameter}.}
\end{figure}

\begin{figure}[htbp]
  \centering
  \includegraphics[width=0.8\textwidth]{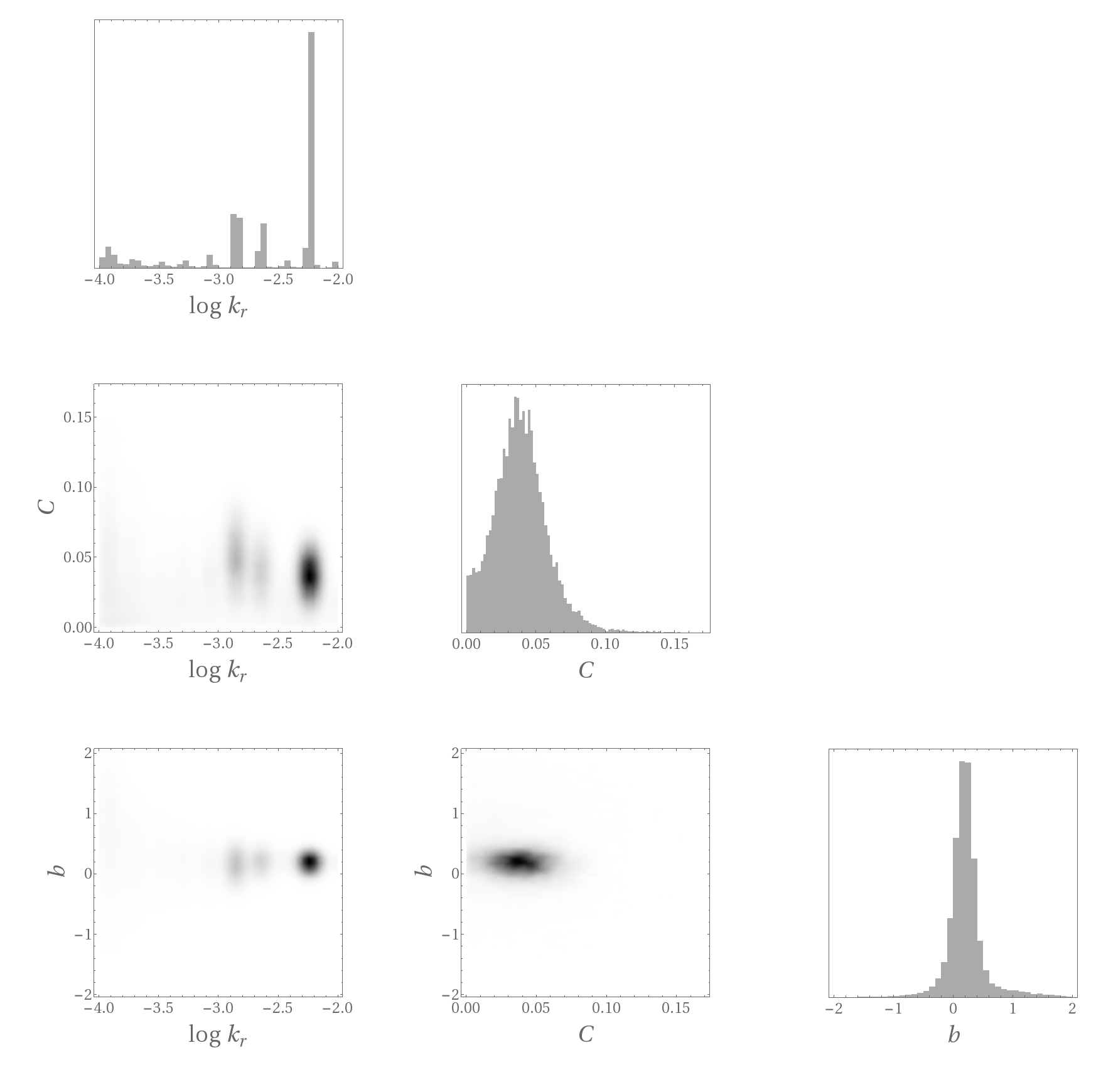}
  \caption{\label{fig:tp1} \small The triangle plot for template \eqref{template3}. }
\end{figure}

The triangle plots of marginalized MCMC samples are given in Fig.~\ref{fig:tp0} for T2 and Fig.~\ref{fig:tp1} for T3.

\begin{table}[t]
  \begin{center}
    \begin{tabular}{ | c || c || c | c |} \hline
              & $\Lambda$CDM & T2     & T3       \\ \hline\hline
$\log k_r$    &            & -2.22     & -2.23       \\ \hline
$C$           &            & 0.0307    & 0.0304      \\
\hline
$b$           &            &           & 0.186       \\
\hline\hline
$\chi^2_{\ell <50}$ & -6.760     & -7.584    & -10.317     \\ \hline
$\chi^2_{\ell \geq 50}$ & 7795.2760  & 7786.590  & 7786.205    \\ \hline
$\chi^2_\mathrm{WP}$    & 2014.305   & 2014.328  & 2014.645    \\ \hline
$\chi^2_\mathrm{total}$ & 9802.821   & 9793.334  & 9790.528    \\ \hline
$\Delta\chi^2_\mathrm{total}$&            & -9.487     & -12.293      \\ \hline
    \end{tabular}
  \end{center}
  \caption{\label{tab:best-chi2-full} \small Best-fit values for templates T2 (\ref{Template_twoparameter}) and T3 (\ref{template3}).}
\end{table}

As mentioned, the parameters in the clock signal determine the location of the associated sharp feature signal. In this full model, we can see that the starting location of the sharp feature indeed falls into the right place as estimated at the end of Sec.~\ref{Sec:Best-fit-parameters}, which coincides with a well-known sharp feature candidate in CMB at around $\ell\sim 20-30$. In the current model, the detailed behavior of this sharp feature around $k=k_0/2$ is either a shallow dip at $k=k_0/2$ or an overall large scale suppression for $k<k_0/2$. For example, see Fig.~\ref{Fig:effect_m0} where we have shown that it is quite natural to generate the overall large scale suppression in the power spectrum for $k<k_0/2$.
Nonetheless, the feature candidate at around $\ell\sim 20-30$ in data could be better fitted by a sharp feature that not only has the overall large scale suppression, but also has a much deeper dip in the primordial spectrum. It would be interesting to see if there are other models with different types of sharp features that can generate such a deep dip.

\section{CMB polarization, non-Gaussianities and LSS}
\label{Sec:Polarization}
\setcounter{equation}{0}

Current and future experiments are producing increasingly precise and complete maps of the entire observable universe. These include maps of different astrophysical objects and spectra, and provide opportunities to constrain or detect beyond-Standard-Model signals through cross correlations between different observables. Primordial feature models are well suited for such opportunities.

Firstly, a single prediction from feature models on a primordial observable typically results in multiple redundant predictions on different cosmological observables. For example, a specific oscillatory pattern predicted by a feature model on the primordial scalar power spectrum results not only in corresponding feature in the temperature power spectrum in CMB, but also correlated features in the polarization power spectrum, the temperature-polarization cross spectrum in CMB, and the matter power spectrum that affects the distribution of large scale structures.

Secondly, feature models themselves typically have multiple correlated predictions on a variety of primordial observables. For example, a feature model typically predicts not only a specific oscillation pattern in the scalar power spectrum, but also correlated patterns in higher order correlation functions such as bispectra and trispectra. Although there are many model-dependent aspects in the properties of these polyspectra, their leading order oscillatory frequencies in the momentum space are strictly correlated \cite{Chen:2006xjb,Chen:2008wn,Chen:2010xka}.

In the following we discuss these aspects in the context of the Standard Clock models.

\subsection{Power spectrum}

\begin{itemize}

\item  {\em Temperature.}
The Standard Clock signals result in fine-structures in the CMB. High resolution temperature data at large multipole moments provided by the Planck experiment is crucial for this purpose. Using the inflationary Standard Clock candidate identified in Sec.~\ref{Sec:MCMC_clock} \& \ref{Sec:MCMC_full} as an example, the Planck temperature data in the frequency channel 143 and 217 GHz are very important because of their high resolutions in the region $\ell >800$. This is the place where the details of the important clock signal of this candidate start to show up.
It will be important to see how the data in this region may be improved in the future analyses, e.g.~by better understanding the foregrounds and increasing the sky coverage (currently $31-37\%$) for these two channels.

\begin{figure}[t]
  \centering
  \begin{tabular}{r}
  \includegraphics[width=0.6\textwidth]{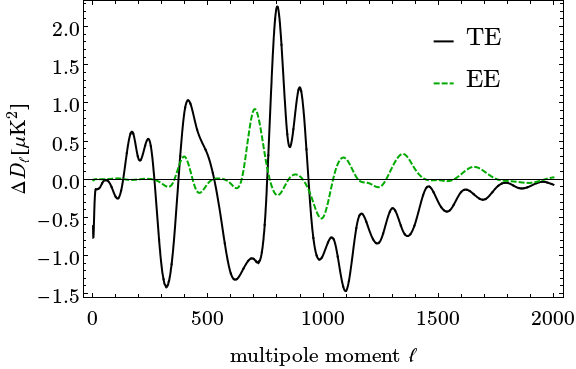}
  \\
  \\
  \includegraphics[width=0.62\textwidth]{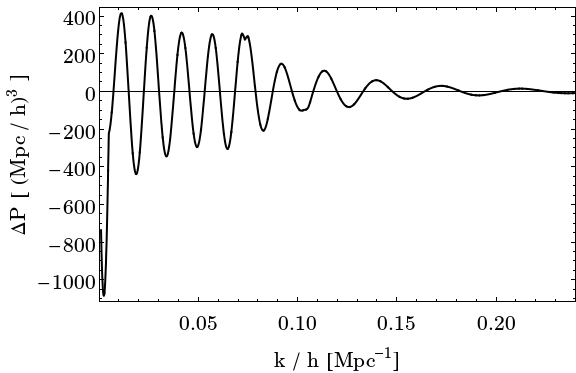}
  \end{tabular}
  \caption{\label{fig:TEEE} \small The corrections to the polarization power spectrum ({\em upper}) and the matter power spectrum ({\em lower}), from the best-fit of  template \eqref{template3} ($T_3$).}
\end{figure}

\begin{figure}[htbp]
  \centering
  \begin{tabular}{r}
  \includegraphics[width=0.625\textwidth]{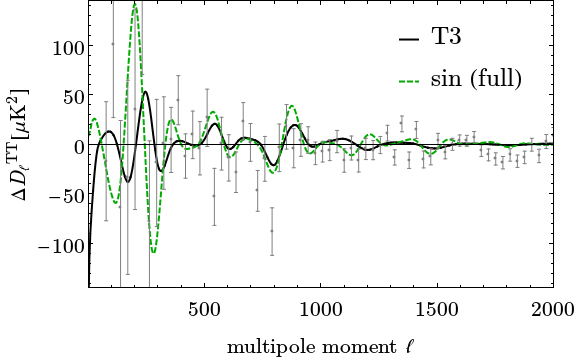}
  \\
  \\
  \includegraphics[width=0.6\textwidth]{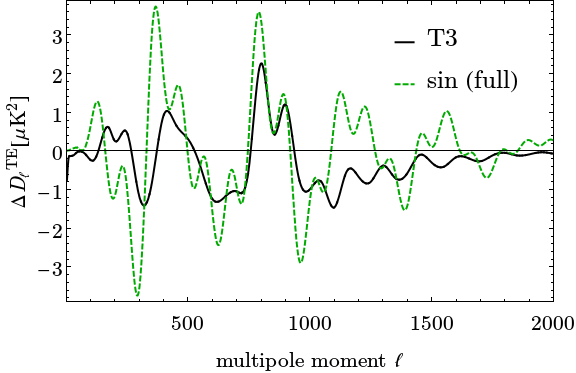}
  \\
  \\
  \includegraphics[width=0.625\textwidth]{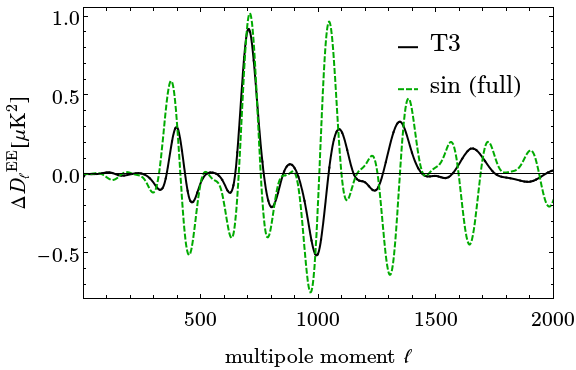}
  \end{tabular}
  \caption{\label{fig:comp_sinclock} \small
  The best-fit model of the full Standard Clock template (\ref{template3}) (T3) versus that of the pure sharp feature signal template (\ref{sharp_sin_template}) (sin(full)), plotted in full range of scales, in terms of the corrections to the TT ({\em upper}), TE ({\em middle}) and EE ({\em lower}) correlations.}
\end{figure}

\item {\em Polarization.}
Primordial scalar power spectrum also contributes to the E-polarization (EE) power spectrum and the temperature-polarization (TE) cross power spectrum. Therefore the Standard Clock features also leave their imprints on these observables. In the soon-to-be-expected new release, the Planck collaboration will release for the first time their high precision polarization data that measures $C^{TE}_\ell$ and $C^{EE}_\ell$ on all scales up to and beyond $\ell \sim 1000$.
CMB polarization has higher foregrounds and noise comparing to temperature data, but it also has advantages that its transfer function is sharper and hence has less damping effect for the primordial oscillations.
In the end we expect that adding the polarization data can potentially decrease the detection threshold for oscillatory signals by $\sim 25\%$ \cite{Fergusson:2014hya}.
Therefore a MCMC analysis with joint temperature and E-polarization data will become important.

In the upper panel of Fig.~\ref{fig:TEEE}, we plot the prediction of the best-fit model from Sec.~\ref{Sec:MCMC_full} on the residuals of the TE and EE angular power spectra. According to the Planck blue book \cite{Planckbluebook}, near $\ell \sim 1000$ with the bin size $\Delta\ell =20$, the $1\sigma$ error for $\CD_\ell^{TE}$ and $\CD_\ell^{EE}$ is $\sim 3 \mu K^2$. From this plot we can see that, for the purpose of the Standard Clock signal, the precision of the TE correlation is comparable to the TT.

In addition, these polarization data will also be valuable in distinguishing the different running behaviors between different feature models.
Of particular interest here is the comparison between the full Standard Clock signal and pure sharp feature signal, as we have discussed in terms of TT correlation in Sec.~\ref{Sec:MCMC_clock}.
In Fig.~\ref{fig:comp_sinclock}, we compare the best-fit model of the full Standard Clock template (\ref{template3}) (T3) with that of the fully extended sharp feature signal template (\ref{sharp_sin_template}) (labelled as ``sin(full)") in terms of the corrections to the TT, TE and EE correlations. For this comparison, the most important property to notice is the phases of the oscillations rather than the envelop behaviors. If we look at the regions $\ell<800$ in all these plots, the phases between T3 and sin(full) are similar in all three figures.\footnote{Note that there are some differences in phases for $\ell\ll 800$. This is due to some small difference between the best-fit frequencies in these two models, which amplifies the phase difference after many oscillations. This difference can be tuned away if we tune the model parameters around the best-fit.} The difference between the Standard Clock and pure sharp feature starts to show up as $\ell>800$.
As we can see, in addition to TT, the TE and EE correlations further help to distinguish these two different oscillatory behaviors.

\item {\em Matter power spectrum.}
The same scalar power spectrum also sets the initial condition for the distribution of large scale structure formation, thereby leaving the imprint of the Standard Clock signal in the matter power spectrum. In the lower panel of Fig.~\ref{fig:TEEE}, we plot the prediction from the best-fit model on the residuals of the matter power spectrum. We expect that the data on the galaxy distribution will be improved significantly in the future due to the large number of galaxy survey experiments. It will be interesting to investigate if the galaxy power spectrum can be used as a good tracer of the underlying matter power spectrum with oscillatory features.

\end{itemize}

\subsection{Non-Gaussianities}

As other feature models, besides generating features in the scalar power spectrum, Standard Clock models also generate highly correlated signals in higher order correlation functions. The most important model-independent property is their running behavior. It has been shown that, at the leading order, these polyspectra all have strictly correlated oscillation frequencies in the momentum space \cite{Chen:2006xjb,Chen:2008wn}. This conclusion can be seen very generally using simple arguments \cite{Chen:2010xka} and has been demonstrated in a variety of examples \cite{Flauger:2010ja,Chen:2010bka,Arroja:2011yu,Adshead:2011jq,Adshead:2012xz,Achucarro:2012fd,Bartolo:2013exa,Martin:2014kja} with model-dependent details.\footnote{For example \cite{Chen:2010bka,Bartolo:2013exa} there may be additional components of non-Gaussianities that have the same frequency but very different $k_i$-dependence, such as the folded shape dependence, comparing to those in (\ref{clock_nonG}) and (\ref{sharp_nonG}).}

For example, as we summarized in Sec.~\ref{Sec:main_properties}, the running behavior of polyspectra for the clock signal is given by
\bea
\sim
\sin \left[ \frac{p^2}{1-p} \Omega \left( \frac{K}{k_r} \right)^{1/p} + {\rm phase} \right] ~,
\label{clock_nonG}
\eea
and the sharp feature signal by
\bea
\sim \sin(K/k_0+{\rm phase}) ~.
\label{sharp_nonG}
\eea
The phases are polyspectra-dependent, but the oscillation frequencies are the same in the $K$-space for the polyspectra. We only need to change the expression for $K$. For example, for power spectrum, $K=k_1+k_2=2k_1$; for bispectrum $K=k_1+k_2+k_3$; and so on.

Searching for these correlated oscillatory signals in polyspectra is an important but challenging task. A method of modal decomposition \cite{Fergusson:2006pr,Fergusson:2009nv,Fergusson:2014gea} has been designed and tested with this goal in mind and is being applied to data analyses. It is also demonstrated \cite{Fergusson:2014hya} that the statistical independency between different correlation functions in the primordial level is also kept in the statistics of the late-time CMB, therefore reinforcing the hope that a combined search for features in polyspectra may significantly increase the statistical significance.

So far we have discussed the model-independent aspects. There are also important model-dependent aspects that need to be worked out. For example, as we did for the power spectrum, a full template of bispectrum that exhibits the details of the connection between the sharp feature and clock signal is desired in data analyses. Also, the quantitative prediction on the amplitude of the bispectrum is also an important question. With all these combined, the full package of predictions on the polyspectra is highly constraining and, in the event of a detection in the model-independent search, would be very informative in probing the details of the models.

\section{Discussions and conclusions}
\label{Sec:Conclusion}
\setcounter{equation}{0}

In this paper we have studied in detail models of Standard Clock as a way to directly measure $a(t)$ of the primordial universe. If observable, this type of signals could provide a direct evidence in distinguishing the inflation scenario from the alternatives. We emphasized the main properties of the Standard Clock signal and the main requirements for the model building. We have motivated and constructed a full model of Standard Clock for the inflation scenario and studied its predictions in different parameter space. This provides a complete example of Standard Clock model that fully realizes several key ingredients proposed previously in \cite{Chen:2011zf,Chen:2011tu}.
We have performed the MCMC analyses in the Planck 2013 temperature data for the clock signal alone, in both the inflation scenario and alternatives. We have also performed the MCMC analyses for the full signal that includes both the sharp feature and clock signal, but only for the inflation scenario. We identified some interesting candidates in the data, although the statistical significance is still marginal. We discussed how such signals may be further tested in different ways using the data from future experiments.

There are many interesting issues that remain to be studied. For example,

$\bullet$ {\em Data analyses.}
This is the most important issue and we have discussed it more extensively in Sec.~\ref{Sec:Polarization}. When new Planck data is available, a joint analysis with the more complete temperature data and the new polarization data is necessary to further assess the significance of the identified candidates and to search for new candidates. A correlated search in both the CMB power spectrum and non-Gaussianities may provide another powerful check. Joint analyses using future high precision data on the distribution of large scale structure may also turn out to be useful.

$\bullet$ {\em Full templates with more variable parameters.}
The current numerical results of the full model can only be accommodated by a full template that has very limited number of variable parameters. So in the MCMC analyses in Sec.~\ref{Sec:MCMC_full}, several other parameters in the model have to be fixed, including the very important frequency parameter $\Omega$. This significantly limits the efficiency of data analyses. It would be interesting to investigate if this limitation can be relaxed in terms of the templates, or if the numerical results may be directly and efficiently fed to the MCMC analyses.

$\bullet$ {\em Non-Gaussianities.}
In this and a previous work \cite{Chen:2014joa} we have constructed and studied a full model of the inflationary Standard Clock. This enables us to provide a full template for the power spectrum analyses. This model also has correlated predictions on the more complicated non-Gaussianities. Similarly to what we did for the power spectrum, although the available templates for the sharp feature and clock signal separately can be used in data analyses, joint full templates for non-Gaussianities are desirable. Also theoretical predictions on the amplitudes and envelops of the non-Gaussianities are very important.

$\bullet$ {\em Full models for alternative scenarios.}
We have only studied full models of the Standard Clock in the inflation scenario. It will be interesting to construct and study full Standard Clock models in alternative scenarios. This study is important because it provides a more complete view of the full signal, especially the important details that connect the sharp feature signal and the clock signal.

$\bullet$ {\em Different full models.}
Requirements for being a Standard Clock model are simple and summarized in the paper. When it comes to data analyses, due to multiple and correlated constraints, model-dependent aspects also become very important.
The two most {\em model-dependent} aspects of the Standard Clock models are the nature of the sharp feature and the nature of the coupling between the clock field and the density-perturbation-source-field.
It is interesting to construct different models of Standard Clock, to check the realization of the shared properties and to provide model-dependent details. For example, as we discussed in the paper, it is interesting to see if there are different types of sharp features that may better model the large dip at around $\ell\sim 20$ in the data and generate the clock signal at the same time; it is also interesting to investigate if the clock signal can be more significant in the large field inflation models.

\medskip
\section*{Acknowledgments}

We would like to thank Christophe Ringeval for collaboration at the initial stage of this work. We would like to thank Nicola Bartolo, George Efstathiou, Razieh Emami, James Fergusson, Steven Gratton, Helge Gruetjen, Alan Guth, Eiichiro Komatsu, Michele Liguori, Sabino Matarrese, and Paul Shellard for helpful discussions.
XC and MHN are supported in part by a NSF grant PHY-1417421. YW is supported by a Starting Grant of the European Research Council (ERC STG grant 279617), and the Stephen Hawking Advanced Fellowship.
The numerical part of this work was undertaken on the COSMOS Shared Memory system at DAMTP, University of Cambridge operated on behalf of the STFC DiRAC HPC Facility. This equipment is funded by BIS National E-infrastructure capital grant ST/J005673/1 and STFC grants ST/H008586/1, ST/K00333X/1. Some analytical calculations were assisted by \verb!MathGR! \cite{Wang:2013mea}.

\appendix

\section{Full perturbative quadratic action}
\setcounter{equation}{0}
\label{App:quadratic_action}

In this appendix we provide some details of the quadratic action for perturbations and try to estimate the strength of each term. As usual we start with the ADM formalism
\begin{align}
\label{ADM}
d S^2 = - N^2 dt^2 + h_{ij} \left(dx^{i} + N^{i} dt \right)\left(dx^{j} + N^{j} dt \right) ~,
\end{align}
in which $N$ and $N_i$ are the lapse function and shift vector, respectively.
The full action is,
\begin{align}
\label{fullaction}
S &= S_m + S_g \nonumber\\
& = \frac{1}{2} \int dt dx^3 \sqrt{h} N \left( R^{(3)} + 2 \CL_m \right) + \frac{1}{2} \int dt dx^3 \sqrt{h} N^{-1} \left( E_{ij}E^{ij} - E^2 \right) ~,
\end{align}
in which the matter Lagrangian is defined by
\ba
\CL_m =
-\frac{1}{2} (\tR + \sigma)^2 g^{\mu\nu} \partial_\mu \theta \partial_\nu \theta - V_{\rm sr}(\theta)
-\frac{1}{2} g^{\mu\nu} \partial_\mu \sigma \partial_\nu \sigma - V_\sigma(\sigma) ~.
\ea
At the background level one can find the following equations of motion
\begin{align}
&
3  \mPl^2 H^2 = \dfrac12 \left(\tilde{R}+\sigma _0\right)^2 \, \dot\theta _0^2+\dfrac12\dot\sigma _0^2+V_{\rm sr}\left(\theta _0\right)+V_{\sigma }\left(\sigma _0\right) ~,
\\
&
\ddot \theta_0 +3 H \dot\theta_0+\dfrac{2 \dot\theta _0 \dot\sigma _0}{(\tilde{R}+\sigma _0)}+\dfrac{V_{\rm sr}'\left(\theta _0\right)}{(\tilde{R}+\sigma _0)^2 }=0 ~,
\\
&
\ddot \sigma _0+3 H \dot\sigma _0+V_{\sigma }'\left(\sigma _0\right)-\left(\tilde{R}+\sigma _0\right) \dot\theta _0^2=0 ~,
\\
&
\epsilon \equiv -\dfrac{\dot H}{H^2} = \dfrac{(\tilde{R}+\sigma _0)^2 \, \dot \theta^2 + \dot \sigma^2}{2 \mPl^2 H^2} ~.
\label{epsilon_definition}
\end{align}

For the perturbations, we use the $\delta\phi$-gauge (or the spatially flat gauge). We perturb the matter sector by $\sigma \to \sigma_0 +\delta \sigma$ and $\theta \to \theta_0 +\delta \theta$, and
$h_{ij}=a^2 \delta_{ij}$ is unperturbed.
We also perturb the lapse function and shift vector as follows
\begin{align}
\label{N and Ni}
N &\simeq 1 + \alpha  ~, \\
N_{i} &\simeq \partial_i \beta  + \widetilde{N}_i.
\end{align}
Since we are interested in the quadratic Lagrangian for scalar mode, we will set the vector modes (e.g. $\widetilde{N}_i$) as well as tensor modes to zero as they will be decoupled from scalar modes at this order. The lapse and shift are non-dynamical parameters and give the following constraints
\begin{align}
\label{constarint equations}
R^{(3)} + 2 \CL_{m} + 2 N \frac{\partial \CL_m}{\partial N} - \frac{1}{N^2} \left(E_{ij}E^{ij} - E^2\right) &=0 ~,
\nonumber\\
\nabla_{i}\left(N^{-1} \left( E^{ij} - h^{ij} E\right) \right) + N \frac{\partial \CL_m}{ \partial N_{j}} & =0 ~,
\end{align}
for which at the linear order we obtain
\ba
\alpha=\frac{\delta \theta  \left(\tilde{R}+\sigma _0\right)^2 \dot\theta _0+\delta \sigma  \dot\sigma _0}{2 H}
\ea
and
\ba
\frac{2H}{a^2}\partial^2 \beta &=&\left( \left( \tilde R + \sigma_{0} \right)^2 \dot{\theta}^2_{0}(t) + \dot{\sigma}^2_{0}(t) - 6 H^2 \right) \alpha
\nonumber \\
&- &  \bigg{(} V'_{sr} \delta \theta + V'_{\sigma} \delta \sigma
+ \left(\tilde  R + \sigma_{0} \right) \dot{\theta}^2_{0}  \delta \sigma + \left( \tilde R + \sigma^2_{0} \right) \dot{\theta}_{0} \delta \dot{\theta} + \dot{\sigma}_{0} \delta \dot{\sigma} \bigg{)} ~.
\ea
Perturbing the full action up to the second order and plugging back the above solutions into the action one can obtain the following Lagrangian
\begin{align}
\CL_2 = &
\frac{a^3}{2} (\tR + \sigma_0)^2
\left[ \dot{\delta\theta}^2 - \frac{1}{a^2} (\partial_i \delta\theta)^2 \right]
+ \frac{a^3}{2} \dot{\delta\sigma}^2 - \frac{a}{2} (\partial_i \delta\sigma)^2
\nonumber \\
&
- \frac{a^3}{2}
\left[V''_{sr}
-(\tR + \sigma_0)^4 \dot \theta^2 (\epsilon-3) +\dfrac{2}{H}(\tR + \sigma_0)^2 \dot \theta V'_{sr}
\right]
 \delta\theta^2
\nonumber\\
&
-\frac{a^3}{2} \left[
V''_\sigma - \dot\theta_0^2
+ \frac{2}{H}\dot\theta_0^2 (\sigma_0 + \tR) \dot\sigma_0
\right]
\delta\sigma^2
\nonumber \\
&
- \frac{a^3}{H} \dot\sigma_0^2 \delta\sigma \dot{\delta\sigma}
\nonumber \\
&
+ \frac{a^3}{H} (\sigma_0+\tR) \dot\theta_0
\left( 2H - (\sigma_0+\tR) \dot\sigma_0 \right)
\delta\sigma \dot{\delta\theta}
\nonumber \\
&
- \frac{a^3}{H} (\sigma_0+\tR)^3 \dot\theta_0^3
\delta\sigma \delta\theta
\nonumber \\
&
- \frac{a^3}{H} (\sigma_0+\tR)^2 \dot\theta_0 \dot\sigma_0
\dot{\delta\sigma} \delta\theta ~,
\label{L2_full}
\end{align}
where the background equations of motion have been used to simplify the coefficients.

In the numerical calculations, all the terms above are used to derive the equations of motions for perturbations. Nonetheless to gain more analytical understanding of the underlying physics, such as several aspects we discuss in Sec.~\ref{Sec:small_field} and \ref{Sec:Large_field}, it is useful to estimate the order of magnitude of each term and find out the most important ones.

To do this we first define
\ba
\epsilon_0 &=& \dfrac{\dot \theta_0^2 \tR^2}{2 \mPl^2 H^2} \simeq 2 \mPl^2 \left( \dfrac{V'_{\rm sr}}{ V_{\rm sr} \tR} \right)^2 ~,
\label{epsilon0_def}
\\
\eta_0 &\simeq &2 \mPl^2 \dfrac{V''_{sr}}{V_{sr}\tR^2} ~,
\ea
where the prime in $V'_{sr}$ denotes the derivative with respect to $\theta_0$. We also approximate $\tR +\sigma_0 \approx \tR$ and $V_{\rm sr} \sim \mpl^2 H^2$.
Notice that the above parameters are different from the slow-roll parameter defined in (\ref{epsilon_definition}). Here we have taken out the contribution from the oscillating massive field, so that both $\epsilon_0$ and $\eta_0$ roughly take the values in the late-time attractor single field solution, and they are both small.

We are interested in two processes, namely the sharp feature transition and the massive field oscillation. The leading order behavior of the background massive field $\sigma_0$ and the perturbations ($\delta\theta$ and $\delta\sigma$) are oscillations. They either oscillate rapidly or make a sharp transition depending on the process, and in both cases the characteristic time scale is $\omega^{-1} \sim m_\sigma^{-1}$. Therefore we can estimate each time derivative on these quantities by a factor of $\omega$. Namely, $\dot\sigma_0 \sim \omega\sigma_0$, $\dot{\delta\theta} \sim \omega \delta\theta$ and $\dot{\delta\sigma} \sim \omega \delta\sigma$.

Now let us use an example to illustrate our estimate. Let us look at the second term in the second table in Table~\ref{Table:Estimate}, labeled as
$(\tR+\sigma_0)^2 \dot\theta_0 \dot\sigma_0/H$. This includes two terms in the Lagrangian (\ref{L2_full}), namely the 2nd term in the 5th line,
$\sim \frac{a^3}{H} (\sigma_0+\tR)^2 \dot\theta_0 \dot\sigma_0 \delta\sigma \dot{\delta\theta}$, and the term in the last line,
$\sim \frac{a^3}{H} (\sigma_0+\tR)^2 \dot\theta_0 \dot\sigma_0 \dot{\delta\sigma} \delta\theta$.
Due to the reasons stated above, both terms can be estimated as
\bea
\left( \frac{\omega^2}{H^3} \tR \dot\theta_0 \sigma_0 \right)
\left( a^3 H^2 \tR\delta\theta \delta\sigma \right) ~.
\eea
In the second bracket we have put in a factor of $\tR$ so that $\tR\delta\theta$ is the canonically normalized field, and a factor of $H^2$ to make the rest of the coefficients in the first bracket dimensionless (after properly restoring the factors of $\mpl$). Using (\ref{epsilon0_def}) we can estimate the order of magnitude of the terms in the first bracket as
\bea
\sqrt{\epsilon_0} \left( \frac{\omega}{H} \right)^2 \left( \frac{\sigma_0}{\mpl} \right) ~,
\eea
which is the value we put in the Table \ref{Table:Estimate}.

All the other terms can be estimated in the similar way.
Note that it is not straightforward to compare the terms in different tables since the behavior of perturbations, $\delta\theta$ and $\delta\sigma$, are quite different. Instead, we compare the terms in each table and pick up the leading ones. In the first table, all three terms can be the leading terms. In the second table, the first two terms can be the leading terms depending on the value of $\sigma_0$ and $\tR$, and we ignore the third term. To compare the terms in the third table, we notice that $(\omega/H)(\sigma_0/\mpl) \ll 1$ because $m_\sigma^2 \sigma_0^2 \ll V_{\rm sr} \sim \mpl^2 H^2$. We also notice that $\omega/H \gg \dot\theta_0/H$. Using these relations, it is clear that only the first term in this table is the dominant term.

\begin{table}
\begin{center}
\begin{tabular}{|c|c|c|c|}
\hline
terms $\propto \delta \theta^2$ & $V''_{sr}$ & $(\tR+\sigma_0)^4 \dot \theta^2 (\epsilon-3)$  & $(\tR+\sigma_0)^2 \dot \theta V'_{sr}/H$
\\
\hline
strength &
$\eta_0$  &
$\epsilon_0 (\epsilon-3)$ &
$\epsilon_0$
\\
\hline
 \end{tabular}

\medskip
\begin{tabular}{|c|c|c|c|}
\hline
terms $\propto \delta \theta \delta \sigma$ & $2 (\tR +\sigma_0) \dot \theta $ & $(\tR+\sigma_0)^2 \dot \theta_0 \dot \sigma_0/H$  &$ (\tR+\sigma_0)^3 \dot \theta_0^3/H$
\\
\hline
strength &
$\sqrt{\epsilon_0} (\dfrac{\omega}{H}) (\dfrac{\tR}{\mPl})^{-1}  $&
$ \sqrt{\epsilon_0} (\dfrac{\omega}{H})^2 (\dfrac{\sigma_0}{\mPl}) $ &
$\epsilon_0^{3/2}  (\dfrac{\tR}{\mPl})^{-1}$
\\
\hline
 \end{tabular}

\medskip
\begin{tabular}{|c|c|c|c|c|}
\hline
terms $\propto \delta \sigma^2$ & $V''_\sigma$ & $ \dot \theta_0^2 $  &$ (\tR+\sigma_0) \dot \theta_0^2 \dot \sigma_0/H$ &
$\dot \sigma_0^2/H$
\\
\hline
strength &
$(\dfrac{\omega}{H})^2$&
$(\dfrac{\dot\theta_0}{H})^2 $ &
$\sqrt{\epsilon_0} (\dfrac{\dot\theta_0}{H}) (\dfrac{\omega}{H}) (\dfrac{\sigma_0}{\mPl})$
&
$(\dfrac{\omega}{H})^3 (\dfrac{\sigma_0}{\mPl})^2$
\\
\hline
 \end{tabular}
 \end{center}
 \caption{\label{Table:Estimate} \small Here we divide terms into three different categories and compare terms in each category by estimating their strength. This determines which terms are important in each category. See the text for details.}
\end{table}

So the terms most important for the two types of feature signals are
\begin{align}
\CL_2 \approx &
\frac{a^3}{2} (\tR + \sigma_0)^2
\left[ \dot{\delta\theta}^2 - \frac{1}{a^2} (\partial_i \delta\theta)^2 \right]
+ \frac{a^3}{2} \dot{\delta\sigma}^2 - \frac{a}{2} (\partial_i \delta\sigma)^2
\nonumber \\
&
- \frac{a^3}{2}
\left[V''_{sr}
-(\tR + \sigma_0)^4 \dot \theta^2 (\epsilon-3) +\dfrac{2}{H}(\tR + \sigma_0)^2 \dot \theta V'_{sr}
\right]
 \delta\theta^2
\nonumber \\
&
-\frac{a^3}{2}
V''_\sigma \delta\sigma^2
\nonumber \\
&
+ \frac{a^3}{H} (\sigma_0+\tR) \dot\theta_0
\left( 2H - (\sigma_0+\tR) \dot\sigma_0 \right)
\delta\sigma \dot{\delta\theta}
\nonumber \\
&
- \frac{a^3}{H} (\sigma_0+\tR)^2 \dot\theta_0 \dot\sigma_0
\dot{\delta\sigma} \delta\theta ~.
\end{align}

\section{Notes on numerical simulations}
\setcounter{equation}{0}
\label{App:notes_numerical}

This Appendix contains some notes on the method of numerical simulations. The full models in Sec.~\ref{Sec:Full_Model} contain several model parameters, all of which can be varied when we try different examples. On the other hand, when we tune parameters, it would be more instructive if we know the effects of these changes on the observables. However, the relations between the model parameters and the final observables are not obvious. For this reason, it is much more efficient if we first find out some rough relations between the observables and the model parameters. Then we can instead tune these observable parameters as the input, and use the rough relations to determine what the model parameters should be.

Here we use the small field model (\ref{V_sigma}) and (\ref{V_smallfield}) as an example.
The model parameters in this model are
\bea
V_{\rm inf} ~, ~~\beta~, ~~V_{\sigma0}~, ~~\sigma_f~, ~~\tR ~,
\label{model_parameters}
\eea
in which $V_{\rm inf}$ is the dominant inflationary potential energy in the slow-roll potential (\ref{V_smallfield}), $\beta$ determines the shape of the slow-roll potential, $V_{\sigma0}$ is the depth of the tachyonic potential dip for the $\sigma$ field, $\sigma_f$ determines the width at the bottom of this dip, $\tR$ is the radius of the turning trajectory in field space.

On the other hand, the following parameters are more directly related to the observables we care about in this paper,
\bea
\epsilon_0 ~, ~~P_{\zeta0}~, ~~\frac{m_\sigma}{H}~, ~~\frac{\sigma_f}{\tR}~, ~~\frac{V_{\sigma0}}{V_{\rm inf}} ~,
\label{observable_parameters}
\eea
where $\epsilon_0$ is the attractor slow-roll parameter, $P_{\zeta0}$ is the leading order power spectrum, $m_\sigma/H$ determines the frequency of the clock $\omega$, the two ratios, $\sigma_f/\tR$ and $m_\sigma/H$, determine the amplitude of the clock signal in the power spectrum through (\ref{Clock_amplitude_estimate}), and $V_{\sigma0}/V_{\rm inf}$ determines the relative depth of the $\sigma$ potential dip to the inflationary energy.

Now we approximately express the model parameters (\ref{model_parameters}) in terms of the observable parameters (\ref{observable_parameters}).
Using
\bea
P_{\zeta0} \approx \frac{H^2}{8\pi^2 \epsilon_0} ~,
\eea
we have
\bea
H\approx 2\sqrt{2} \pi \sqrt{\epsilon_0 P_{\zeta0}} ~.
\label{H_approx}
\eea
Note that $H^2 \approx V_{\rm inf}/3$, we get
\bea
V_{\rm inf} \approx 24\pi^2 \epsilon_0 P_{\zeta0} ~.
\label{V_inf_approx}
\eea

From the shape at the tip of the potential dip (\ref{V_sigma}), we get
$m_\sigma \approx \sqrt{2 V_{\sigma0}}/\sigma_f$. This relation can be used to express $\tR$ in terms of the observable parameters,
\bea
\tR = \frac{\sigma_f}{\sigma_f/\tR} \approx
\sqrt{6} \frac{\sqrt{V_{\sigma0}/V_{\rm inf}}}{(m_\sigma/H)(\sigma_f/\tR)} ~.
\label{tR_approx}
\eea

Using the attractor background equation of motion
\bea
\frac{V'_{\rm sr}(\theta)}{\tR} \approx -3H\tR \dot\theta_0
\eea
and the relation
\bea
\epsilon_0 \approx \frac{\tR^2 \dot\theta_0^2}{2H^2} ~,
\eea
we find the expression for the shape of the slow-roll potential $\beta$
\bea
\beta \approx 3 \sqrt{2} \sqrt{\epsilon_0} H^2 \tR
\approx 48\sqrt{3} \pi^2 \epsilon_0^{3/2} P_\zeta
\frac{\sqrt{V_{\sigma0}/V_{\rm inf}}}{(m_\sigma/H)(\sigma_f/\tR)} ~.
\label{beta_approx}
\eea

Finally, the expressions for $V_{\sigma0}$ and $m_\sigma$ are simple to get,
\begin{align}
V_{\sigma0} &\approx \frac{V_{\sigma0}}{V_{\rm inf}} 3H^2
\approx 24\pi^2 \epsilon_0 P_{\zeta0} \frac{V_{\sigma0}}{V_{\rm inf}} ~,
\label{V_sigma0_approx}
\\
\sigma_f &= \frac{\sigma_f}{\tR} \tR \approx \sqrt{6}
\frac{\sqrt{V_{\sigma0}/V_{\rm inf}}}{m_\sigma/H} ~.
\label{sigma_f_approx}
\end{align}

In summary, (\ref{V_inf_approx}), (\ref{tR_approx}), (\ref{beta_approx}), (\ref{V_sigma0_approx}) and (\ref{sigma_f_approx}) are the relations that we use when adjusting the parameters in the numerical codes. The relations for the large field inflation model (\ref{V_largefield}) can be obtained in a similar way.

\end{spacing}

\end{document}